\documentclass[prd,amsmath,amssymb,showpacs,superscriptaddress,nofootinbib]{revtex4}

\usepackage{amssymb}
\usepackage{multirow}
\usepackage{graphicx,epsfig}% Include figure files
\usepackage{dcolumn}% Align table columns on decimal point
\usepackage{bm}% bold math
\usepackage[dvipdfm,CJKbookmarks=true,unicode,colorlinks,linkcolor=blue,anchorcolor=blue,citecolor=blue,pdfborder={0 0 0}]{hyperref}
\usepackage{amsfonts}
\usepackage{keyval,graphicx}
\usepackage{textcomp,wasysym}

\begin{document}
%\linenumbers
%\baselineskip=20pt
\newcommand{\cinst}[2]{$^{\mathrm{#1}}$~#2\par}
\newcommand{\crefi}[1]{$^{\mathrm{#1}}$}
\newcommand{\crefii}[2]{$^{\mathrm{#1,#2}}$}
\newcommand{\crefiii}[3]{$^{\mathrm{#1,#2,#3}}$}
\newcommand{\HRule}{\rule{0.5\linewidth}{0.5mm}}
\newcommand{\br}[1]{\mathcal{B}#1}
\newcommand{\el}[1]{\mathcal{L}#1}
\newcommand{\ef}[1]{\mathcal{F}#1}

\parskip=3pt plus 1pt minus 1pt

\title{\boldmath  Partial wave analysis of $J/\psi \to \gamma \eta \eta$}
%===========================================================================
% Author
%===========================================================================
\author{
\small
M.~Ablikim$^{1}$, M.~N.~Achasov$^{6}$, O.~Albayrak$^{3}$, D.~J.~Ambrose$^{39}$, F.~F.~An$^{1}$, Q.~An$^{40}$, J.~Z.~Bai$^{1}$, R.~Baldini Ferroli$^{17A}$, Y.~Ban$^{26}$, J.~Becker$^{2}$
% ...
\\
\vspace{0.2cm}
(BESIII Collaboration)\\
\vspace{0.2cm} {\it
$^{1}$ Institute of High Energy Physics, Beijing 100049, People's Republic of China\\
$^{2}$ Bochum Ruhr-University, D-44780 Bochum, Germany\\
$^{3}$ Carnegie Mellon University, Pittsburgh, Pennsylvania 15213, USA\\
% ...
$^{6}$ G.I. Budker Institute of Nuclear Physics SB RAS (BINP), Novosibirsk 630090, Russia\\
% ...
$^{17}$ (A)INFN Laboratori Nazionali di Frascati, I-00044, Frascati, Italy; (B)INFN and University of Perugia, I-06100, Perugia, Italy\\
% ...
$^{26}$ Peking University, Beijing 100871, People's Republic of China\\
% ...
$^{39}$ University of Rochester, Rochester, New York 14627, USA\\
$^{40}$ University of Science and Technology of China, Hefei 230026, People's Republic of China\\
% ...
}}

\vspace{0.4cm}

%\date{\today}

\begin{abstract}

  Based on a sample of $2.25\times 10^{8}$ $J/\psi$ events collected
  with the BESIII detector at BEPCII, a full partial wave analysis
  on $J/\psi\to\gamma\eta\eta$ was performed using the relativistic
  covariant tensor amplitude method. The results show that the
  dominant $0^{++}$ and $2^{++}$ components are from
  the $f_0(1710)$, $f_0(2100)$, $f_0(1500)$, $f_2'(1525)$, $f_2(1810)$ and $f_2(2340)$.
  The resonance parameters and branching fractions are also presented. \\

\end{abstract}


\maketitle

%%%%%%%%%%%%%%%%%%%%%%%%%%%%%%%%%%%%%%%%%%%%%%%%%%%%%%%%%%%%%%%%%%%%%%%%%%%%%%%%%%%%%%%%%%%%%%%%%%%%
\section{Introduction}

Our present understanding of the strong interaction is based on a
non-Abelian gauge field theory, Quantum Chromodynamics
(QCD), which describes the interactions of quarks
and gluons; it also predicts the existence of new types of hadrons
with explicit gluonic degrees of freedom (eg. glueballs, hybrids and
multi-quarks)~\cite{Close:1987er,Godfrey:1998pd,Amsler:2004ps,Klempt:2007cp,Crede:2008vw}.
These unconventional states, if they exist, will
enrich the meson spectroscopy greatly and shed light on the dynamics
of QCD. According to lattice QCD predictions~\cite{Chen:2005mg,Gregory:2012hu}, the
lowest mass glueball with $J^{PC}=0^{++}$ is in the mass region from $1.5$
to $1.7$~GeV/$c^2$.  However, the mixing of the pure glueball with
nearby $q \bar q$ nonet mesons makes the identification of the
glueballs difficult in both experiment and theory.
Radiative $J/\psi$ decay is a gluon-rich process and
has long been regarded as one of the most promising hunting grounds for
glueballs. In particular, for a $J/\psi$ radiative decay to two
pseudoscalar mesons, it offers a very clean laboratory to search for
scalar and tensor glueballs because only intermediate states with
$J^{PC}=even^{++}$ are possible.  An early study of $J/\psi \to
\gamma \eta \eta$ was made by the Crystal Ball
Collaboration~\cite{Edwards:1981ex} with the first observation of
$f_0(1710)$, but the study suffered from low statistics.

In this paper, the results of partial wave analysis (PWA) on
$J/\psi\to\gamma\eta\eta$ are presented based on a sample of
$2.25\times 10^{8}$ $J/\psi$ events~\cite{jpsinumber} collected with
the Beijing Spectrometer (BESIII) located at the upgraded Beijing
Electron and Positron Collider (BEPCII)~\cite{bepc2}.

%%%%%%%%%%%%%%%%%%%%%%%%%%%%%%%%%%%%%%%%%%%%%%%%%%%%%%%%%%%%%%%%%%%%%%%%%%%%%%%%%%%%%%%%%%%%%%%%%%%%

\section{Detector and Monte Carlo simulation}

The BESIII detector, described in detail in Ref.~\cite{bes3dect}, has
an effective geometrical acceptance of 93\% of 4$\pi$. It contains a
small cell helium-based main drift chamber (MDC) which provides
momentum measurements of charged particles; a time-of-flight system
(TOF) based on plastic scintillator which helps to identify charged
particles; an electromagnetic calorimeter (EMC) made of CsI (Tl)
crystals which is used to measure the energies of photons and provide
trigger signals; and a muon system (MUC) made of Resistive Plate
Chambers (RPC). The momentum resolution of charged particles is
$0.5$\% at 1~GeV$/c$ in a 1~Tesla magnetic field. The energy loss
($d\mathrm{E}/dx$) measurement provided by the MDC has a resolution
better than 6\% for electrons from Bhabha scattering. The photon
energy resolution can reach $2.5$\% ($5$\%) at 1.0~GeV in the
barrel (endcaps) of the EMC. And the time resolution of TOF is $80$~ps
in the barrel and $110$~ps in the endcaps.

Monte Carlo~(MC) simulated events are used to determine the detection
efficiency, optimize the selection criteria, and study the possible
backgrounds. The simulation of the BESIII detector, where the
interactions of the particles with the detector material are
simulated, is {\sc geant4}~\cite{Agostinelli:2002hh} based. The
$J/\psi$ resonance is produced with
\textsc{kkmc}~\cite{kkmc2000,kkmc2001}, while the subsequent decays
are generated with {\sc EvtGen}~\cite{Ping2008}.  The study of the
background is based on a MC sample of $2.25\times 10^8$ $J/\psi$
inclusive decays which are generated with known branching fractions
taken from the Particle Data Group (PDG)~\cite{pdg2012}, or with {\sc
lundcharm}~\cite{Lundcharm} for the unmeasured decays.

%%%%%%%%%%%%%%%%%%%%%%%%%%%%%%%%%%%%%%%%%%%%%%%%%%%%%%%%%%%%%%%%%%%%%%%%%%%%%%%%%%%%%%%%%%%%%%%%%%%%
\section{Event selection}
In this analysis, the $\eta$ meson is detected in its $\gamma\gamma$
decay. Each candidate event is required to have five or six good
photons and no charged tracks.  The photon candidates are selected
from the showers in the EMC with deposited energy in the EMC barrel
region ($|\cos\theta|<0.8$) and EMC endcap region
($0.86<|\cos\theta|<0.92$) greater than 25~MeV and 50~MeV,
respectively. The energy deposit in nearby TOF counters is included to
improve the reconstruction efficiency and energy resolution.

To suppress the background events with $\pi^0$
(eg. $J/\psi\to\gamma\pi^0\pi^0$), the events that satisfy
$|M_{\gamma\gamma}-m_{\pi^0}| < 0.015$~GeV/$c^{2}$ are removed, where
$M_{\gamma\gamma}$ is the invariant mass of any pair of photon
candidates and $m_{\pi^0}$ is the nominal $\pi^0$ mass~\cite{pdg2012}.
Then a four-constraint kinematic fit (4C), imposing energy-momentum
conservation, is performed under the $J/\psi\to 5\gamma$ hypothesis to
reduce background events and improve the mass resolution, and
$\chi^2_{4C}$ is required to be less than 50. If the number of
selected photons is larger than five, the fit is repeated using all
permutation of the photons, and the combination with the smallest
$\chi^2_{4C}$ is selected.

\begin{figure*}[htbp]
   \vskip -0.1cm
   \centering
   {\includegraphics[width=5.cm,height=5.cm]{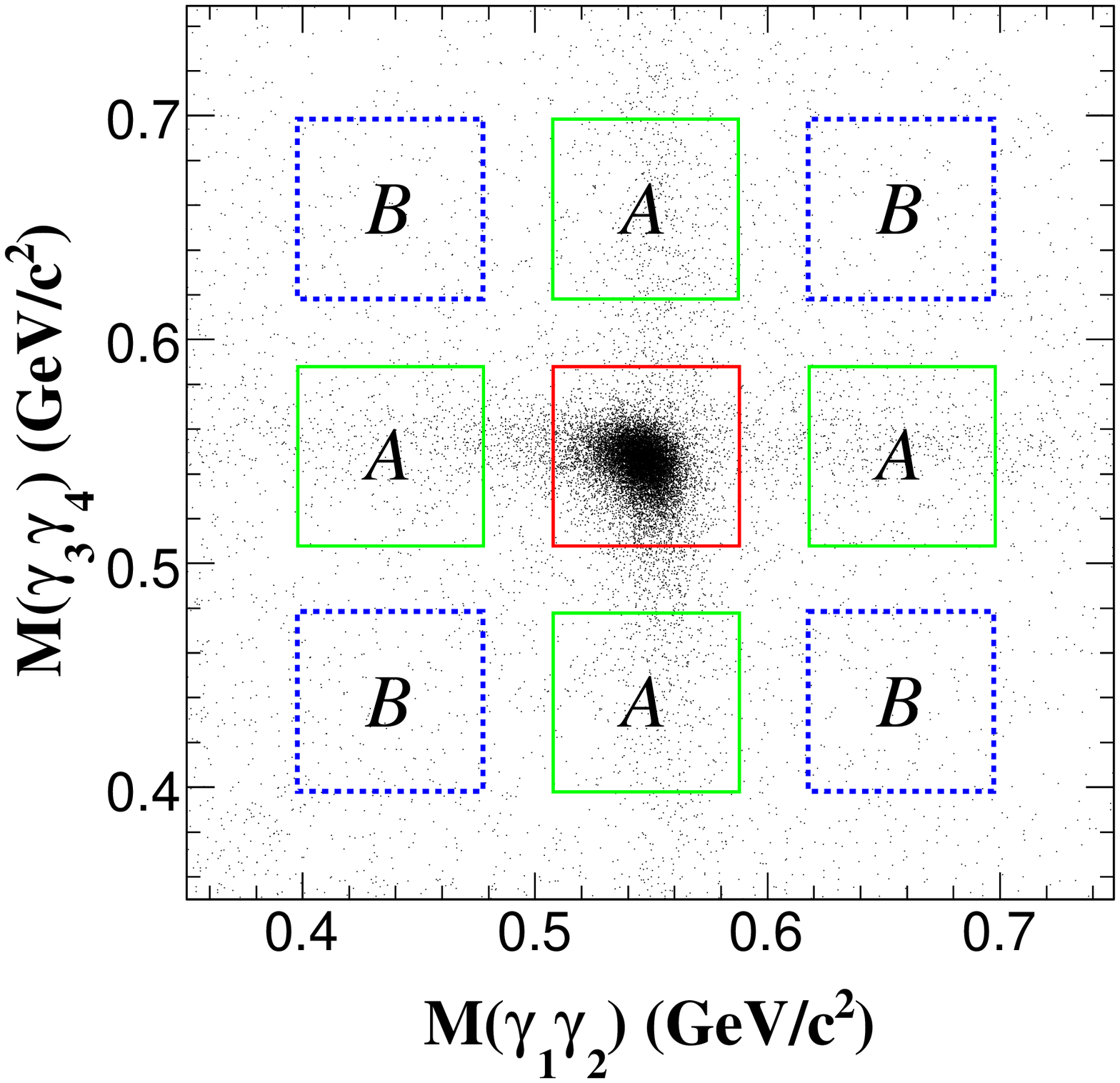}
    \put(-140,5){(a)}}
   {\includegraphics[width=5.cm,height=5.cm]{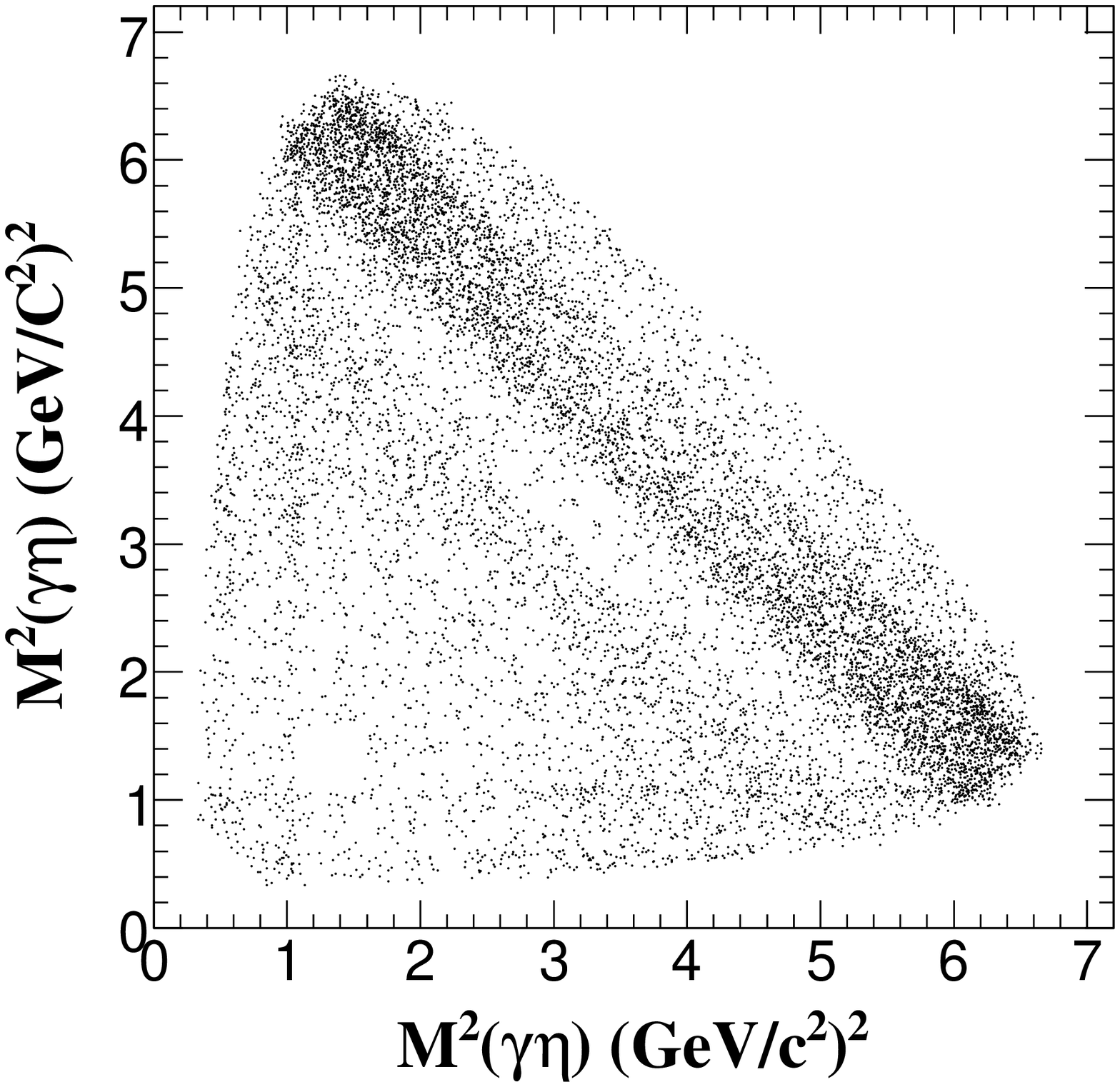}
    \put(-140,5){(b)}}
   {\includegraphics[width=5.cm,height=5.cm]{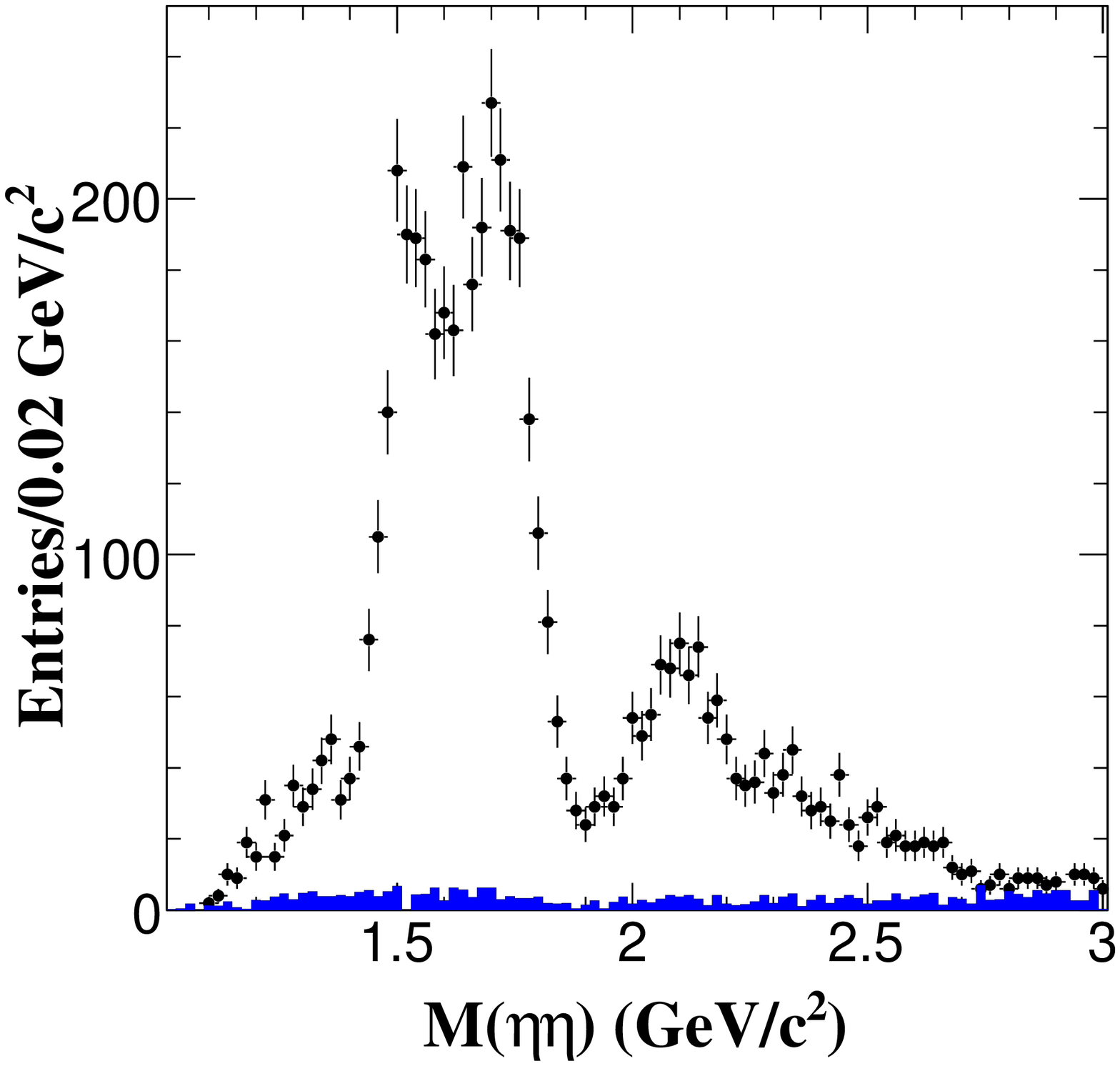}
    \put(-140,5){(c)}}
  \caption{ (a) The scatter plot of $M_{\gamma_1\gamma_2}$ versus
    $M_{\gamma_3 \gamma_4}$ after selecting candidates with the
    minimum $\delta$ (two entries/event).  The two-dimensional $\eta$
    sidebands are framed in regions A and B. (b) Dalitz plot (two
    entries/event), before $\phi\to\gamma\eta$ veto. (c) The invariant mass spectrum of
    $\eta\eta$. The dots with error bars are data, and the shaded
    histogram is background estimated from $\eta$ sidebands.}
   \label{fig:data}
   \vskip -0.5cm
\end{figure*}

To distinguish the photons from $\eta$ decays, a variable $\delta$, defined as
$\delta=\sqrt{(M_{\gamma_{1}\gamma_{2}}-m_{\eta})^{2}+(M_{\gamma_{3}\gamma_{4}}-m_{\eta})^{2}}$,
is introduced, and the combination with the minimum value of $\delta$
is chosen.  The scatter plot of the invariant mass of one $\eta$
candidate versus the other is shown in Fig.~\ref{fig:data}~(a), where
the decay $J/\psi\to\gamma\eta\eta$ is clear.  In order to select a
clean sample, both $M_{\gamma_1\gamma_2}$ and $M_{\gamma_3\gamma_4}$
are required to be in the $\eta$ mass region,
$\left|{M_{\gamma_1\gamma_2}}(M_{\gamma_3\gamma_4}) -m_{\eta} \right|
< 0.04$~GeV/$c^2$, with $m_{\eta}$ the nominal $\eta$
mass~\cite{pdg2012}. The mass resolution for $m_{\eta}$ is about 10~MeV/$c^{2}$.

MC study shows that after the above selection, about 5.3\% of events
have a mis-combination of photons, which mainly occurs between the
radiative photon and one photon from an $\eta$. Therefore, candidate
event must have only one combination with $\delta<0.05$ GeV/$c^2$ to
remove these events, which reduces the fraction of events with a
mis-combination of photons to be 0.8\%.

After that, clear diagonal bands, which correspond to the structures
observed in the $\eta \eta$ invariant mass spectrum, can be
seen in the Dalitz plot for the selected $J/\psi\to\gamma\eta\eta$
candidate events (Fig.~\ref{fig:data}~(b)). A further requirement on
$M_{\gamma\eta}$, $|M_{\gamma\eta}-m_{\phi}|>$30~MeV/$c^{2}$, is used
to reject background events from $J/\psi\to\phi\eta
(\phi\to\gamma\eta$). Fig.~\ref{fig:data}~(c) shows the $\eta\eta$
invariant mass spectrum of the surviving 5460 events after the event
selection.

Potential background events have been studied using both MC and
data samples. The non-$\eta$ events are determined from the two dimensional
$\eta$ mass-sidebands,
0.07~GeV/$c^{2}<\left|M_{\gamma\gamma}-m_{\eta}\right|<$
0.15~GeV/$c^{2}$, which are defined as frame A and B in
Fig.~\ref{fig:data}~(a). The shaded histogram in Fig.~\ref{fig:data}~(c)
shows the sideband events normalized according to  $J/\psi\to 5\gamma$ phase space MC.
In addition, the background events are studied with a MC sample of 225
million $J/\psi$ inclusive events, and the main background events are
found to be from $J/\psi\rightarrow\gamma\eta\pi^0\pi^0$ and
$\gamma\pi^0\eta$. In this paper, the background events estimated from
$\eta$ mass-sidebands, corresponding to a background level of 6\%, are
used in the partial wave analysis below.

\section{Partial Wave Analysis}
%%%%%%%%%%%%%%%%%%%%%%%%%%%%%%%%%%%%%%%%%%%%%%%%%%%%%%%%%%%%%%%%%%%%%%%%%%%%%%%%%%%%%%%%%%%%%%%%%%%%%%%%%%%%%%%%%%%%%%%%%%%%%%%%%%%%%%%%%%%

\subsection{Analysis method}

With GPUPWA, a Partial Wave Analysis framework harnessing GPU parallel
computing~\cite{GPUPWA}, a PWA was performed to disentangle the
structures present in $J/\psi\to\gamma\eta\eta$ decays.  The quasi
two-body decay amplitudes in the sequential decay process
$J/\psi\to\gamma X, X\to\eta\eta$ are constructed using covariant
tensor amplitudes described in Ref.~\cite{Zou:2002ar}. For $J/\psi$
radiative decay to mesons, the general form for covariant tensor
amplitude is:
\begin {equation}
A=\psi_{\mu}(m_{1})e_{\nu}^{*}(m_{2})A^{\mu\nu}=\psi_{\mu}(m_{1})e_{\nu}^{*}(m_{2})\Sigma_{i}\Lambda_{i}U_{i}^{\mu\nu},
\end {equation}
where $\psi_{\mu}(m_{1})$ is the $J/\psi$ polarization four-vector,
$e_{\nu}(m_{2})$ is the polarization vector of the photon and
$U_{i}^{\mu\nu}$ is the partial wave amplitude with coupling strength
determined by a complex parameter $\Lambda_{i}$. The partial wave
amplitudes $U_{i}$ for the intermediate states used in the analysis
are constructed with the four-momenta of daughter particles, and their
specific expressions are given in Ref.~\cite{Zou:2002ar}.

For an intermediate resonance, the corresponding Breit-Wigner
propagator is described by a function:
\begin {equation}
BW(s)=\frac{1}{M^{2}-s-iM\Gamma},
\end {equation}
where s is the invariant mass-squared of daughter particles, and $M$
and $\Gamma$ are the mass and width of the intermediate resonance.

The relative magnitudes and phases of the amplitudes are determined by
an unbinned maximum likelihood fit.
The resonance parameters are optimized by the scan method: repeating
the fits iteratively with various masses and widths until the
optimized likelihood values converge. For the tensor states, the
relative phases between three amplitudes for a certain resonance are
theoretically expected to be very small~\cite{tensorphases}; therefore
the relative phases are set to zero in the fit so as to constrain the
intensities further.

The basis of the likelihood fitting is that a hypothesized probability
density function (PDF) would produce the data set under
consideration. The probability to observe the event characterized by
the measurement $\xi$ is :
\begin {equation}
P(\xi)=\frac{\omega(\xi)\epsilon(\xi)}{\int d\xi\omega(\xi)\epsilon(\xi)},
\end {equation}
where $\epsilon(\xi)$ is the detection efficiency and
$\omega(\xi)\equiv\frac{d\sigma}{d\Phi}$ is the differential cross
section, and $d\Phi$ is the standard element of phase space. The full
differential cross section is:
\begin {equation}
\small
\frac{d\sigma}{d\Phi}=|\sum_{j}{A_{j}}|^{2}=|A(V\eta)+A(0^{++})+A(2^{++})+A(4^{++})+...|^{2},
\end {equation}

where $A(J^{PC})$ is the full amplitude for all resonances whose
spin-parity are $J^{PC}$, and $A(V \eta)$ is the contribution of
the sequential decay processes such as
$J/\psi\to\phi\eta\to\gamma\eta\eta$. $\int
d\xi\omega(\xi)\epsilon(\xi)\equiv\sigma'$ is the measured total cross
section.

The joint probability density for observing the $N$ events in the data
sample is:
\begin {equation}
\el=\prod\limits_{i=1}^{N}P(\xi_{i})=\prod\limits_{i=1}^{N}\frac{(\frac{d\sigma}{d\Phi})_{i}\epsilon(\xi_{i})}{\sigma'}.
\end {equation}

For the technical reasons, rather than maximizing $\el$, $\cal S$=-$\ln\el$ is minimized, i.e.,
\begin {equation}
\ln\el=\sum_{i=1}^{N}\ln(\frac{(\frac{d\sigma}{d\Phi})_{i}}{\sigma'})+\sum_{i=1}^{N}\ln\epsilon(\xi_{i}),
\end {equation}
for a given data set, the second term is a constant and has no impact
on the determination of the parameters of the amplitudes or on the
relative changes of $\cal S$ values. So, for the fitting, $\ln\el$ defined
as:
\begin {equation}
\ln\el=\sum_{i=1}^{N}\ln(\frac{(\frac{d\sigma}{d\Phi})_{i}}{\sigma'})=\sum_{i=1}^{N}\ln{(\frac{d\sigma}{d\Phi})_{i}}-N\ln{\sigma'},
\end {equation}
is used. The free parameters are optimized by
FUMILI~\cite{Dymov:1998zu}. The measured total cross section $\sigma'$
is evaluated using MC techniques. Namely, a MC sample of $N_{gen}$ is
generated with signal events that are distributed uniformly in phase
space. These events are subjected to our selection criteria and yield
a sample of $N_{acc}$ accepted events.  The normalization integral is
computed as:
\begin {equation}
\int {d\xi}\omega(\xi)\epsilon(\xi)=\sigma'\to\frac{1}{N_{acc}}\sum_{k}^{N_{acc}}(\frac{d\sigma}{d\Phi})_{k}.
\end {equation}

The background contribution is estimated with $\eta$ sidebands. In the
log likelihood calculation, the likelihood value of $\eta$ sidebands
events are given negative weights, and are removed from data since the
log likelihood value of data is the sum of the log likelihood values
of signal and background events, i.e.,
\begin {equation}
\cal S=-(\ln\el_{Data} - \ln\el_{BG}).
\end {equation}

The number of the fitted events $N_{X}$ for an intermediate resonance
X, which has $N_{W}$ independent partial wave amplitudes $A_{i}$, is
defined as:
\begin{equation}
 N_{X}=\frac{\sigma_{X}}{\sigma'}\cdot N',
\end{equation}
where $N^{'}$ is the number of selected events after background
subtraction, and
\begin{equation}
\sigma_{X}=\frac{1}{N_{acc}}\sum_{k}^{N_{acc}}|\sum_{j}^{N_{W}}(A_{j})_{k}|^{2},
\end{equation}
is the measured cross section of the resonance X and is calculated
with the same MC sample as the measured total cross section $\sigma'$.

The branching ratio of $J/\psi\to\gamma X,X \to\eta\eta$ is calculated with:
\begin {equation}
\br(J/\psi\rightarrow\gamma X\rightarrow\gamma\eta\eta)=\frac{N_{X}}{N_{J/\psi}\cdot\varepsilon_{X}
\cdot \br^{2}_{\eta\rightarrow\gamma\gamma}},
\end {equation}
where the detection efficiency $\varepsilon_{X}$ is obtained by
the partial wave amplitude weighted MC sample,
\begin{equation}
\varepsilon_{X}=\frac{\sigma_{X}}{\sigma_{X}^{gen}}=\frac{\sum_{k}^{N_{acc}}|\sum_{j}^{N_{W}}(A_{j})_{k}|^{2}}{\sum_{i}^{N_{gen}}|\sum_{j}^{N_{W}}(A_{j})_{i}|^{2}}.
\end{equation}

The statistical errors for masses, widths and branching ratios in a PWA are
defined as one standard deviation from the optimized results, which
corresponds to a change, 0.5, of the log likelihood value for a
specific parameter. In this analysis, the changes of log likelihood
value and the number of free parameters in the fit with or without a
resonance are used to evaluate the statistical significance of this resonance.

%%%%%%%%%%%%%%%%%%%%%%%%%%%%%%%%%%%%%%%%%%%%%%%%%%%%%%%%%%%%%%%%%%%%%%%%%%%%%%%%%%%%%%%%%%%%%%%%%%%%%%%%%%%%%%%%%%%%%%%%%%%%%%%%%%%%%%%%%%%

\subsection{PWA results}
In this analysis, all possible combinations of $0^{++}$, $2^{++}$,
$4^{++}$ resonances listed in the PDG summary table~\cite{22states} are
evaluated, and the fitted components with statistical significance
larger than 5.0$\sigma$
are kept as the basic solution. The contribution from $4^{++}$
($f_4(2050)$) with a statistical significance of 0.4$\sigma$ is
ignored.  There are six resonances, $f_{0}(1500)$, $f_{0}(1710)$,
$f_{0}(2100)$, $f_{2}^{'}(1525)$, $f_{2}(1810)$, $f_{2}(2340)$, as
well as $0^{++}$ phase space and $J/\psi\to\phi\eta$ included in the
basic solution.  Although most of the $J/\psi\to\phi\eta$ events have
been rejected by the above $\phi$ mass window requirement,
$J/\psi\to\phi\eta$ is included in the PWA to evaluate its impact from
the interference between the tail of $\phi$ and other components from
$J/\psi\to\gamma X (X\rightarrow \eta\eta)$. The masses and widths of
the resonances, branching ratios of $J/\psi$ radiative decaying to X
and the statistical significances are summarized in Table~\ref{mwb}.

%%%%%%%%%%%%%%%%%%%%%%%%%%%%%%%%%%%%%%%%%%%%%%%%%%%%%%%%%%%%%%%%%%%%%%%%%%%%%%%%%%%%%%%%%%%%%%%%%%%%%%%%%%%%%%%%%%%%%%%%%%%%%%%%%%%%%%%%%%%
%% Table I
%%%%%%%%%%%%%%%%%%%%%%%%%%%%%%%%%%%%%%%%%%%%%%%%%%%%%%%%%%%%%%%%%%%%%%%%%%%%%%%%%%%%%%%%%%%%%%%%%%%%%%%%%%%%%%%%%%%%%%%%%%%%%%%%%%%%%%%%%%%
\begin {table*}[htb]
\begin {center}
\caption {Summary of the PWA results, including the masses and widths for resonances, branching ratios of
$J/\psi\to\gamma$X, as well as the significance. The first errors are statistical and the second ones are systematic.
The statistic significances here are obtained according to the changes of the log likelihood.}
\begin {tabular}{ccccc}
\hline\hline Resonance  &Mass(MeV/$c^{2}$) &Width(MeV/$c^{2}$)  &$\br{(J/\psi\to\gamma X\to\gamma \eta\eta)}$ &Significance\\ \hline

$f_{0}(1500)$  &1468$^{+14+23}_{-15-74}$  &136$^{+41+28}_{-26-100}$    &$(1.65^{+0.26+0.51}_{-0.31-1.40})\times10^{-5}$  &8.2~$\sigma$   \\%\hline

$f_{0}(1710)$  &1759$\pm6^{+14}_{-25}$    &172$\pm10^{+32}_{-16}$      &$(2.35^{+0.13+1.24}_{-0.11-0.74})\times10^{-4}$  &25.0~$\sigma$  \\%\hline

$f_{0}(2100)$  &2081$\pm13^{+24}_{-36}$   &273$^{+27+70}_{-24-23}$     &$(1.13^{+0.09+0.64}_{-0.10-0.28})\times10^{-4}$  &13.9~$\sigma$  \\%\hline

$f_{2}^{'}(1525)$  &1513$\pm5^{+4}_{-10}$  &75$^{+12+16}_{-10-8}$      &$(3.42^{+0.43+1.37}_{-0.51-1.30})\times10^{-5}$  &11.0~$\sigma$  \\%\hline

$f_{2}(1810)$  &1822$^{+29+66}_{-24-57}$  &229$^{+52+88}_{-42-155}$    &$(5.40^{+0.60+3.42}_{-0.67-2.35})\times10^{-5}$  &6.4~$\sigma$   \\%\hline

$f_{2}(2340)$  &2362$^{+31+140}_{-30-63}$  &334$^{+62+165}_{-54-100}$  &$(5.60^{+0.62+2.37}_{-0.65-2.07})\times10^{-5}$  &7.6~$\sigma$   \\\hline \hline

\end {tabular}
\label{mwb}
\end {center}
\end {table*}

\begin{figure*}[htbp]
   \vskip -0.1cm
   \centering
   {\includegraphics[height=5.cm]{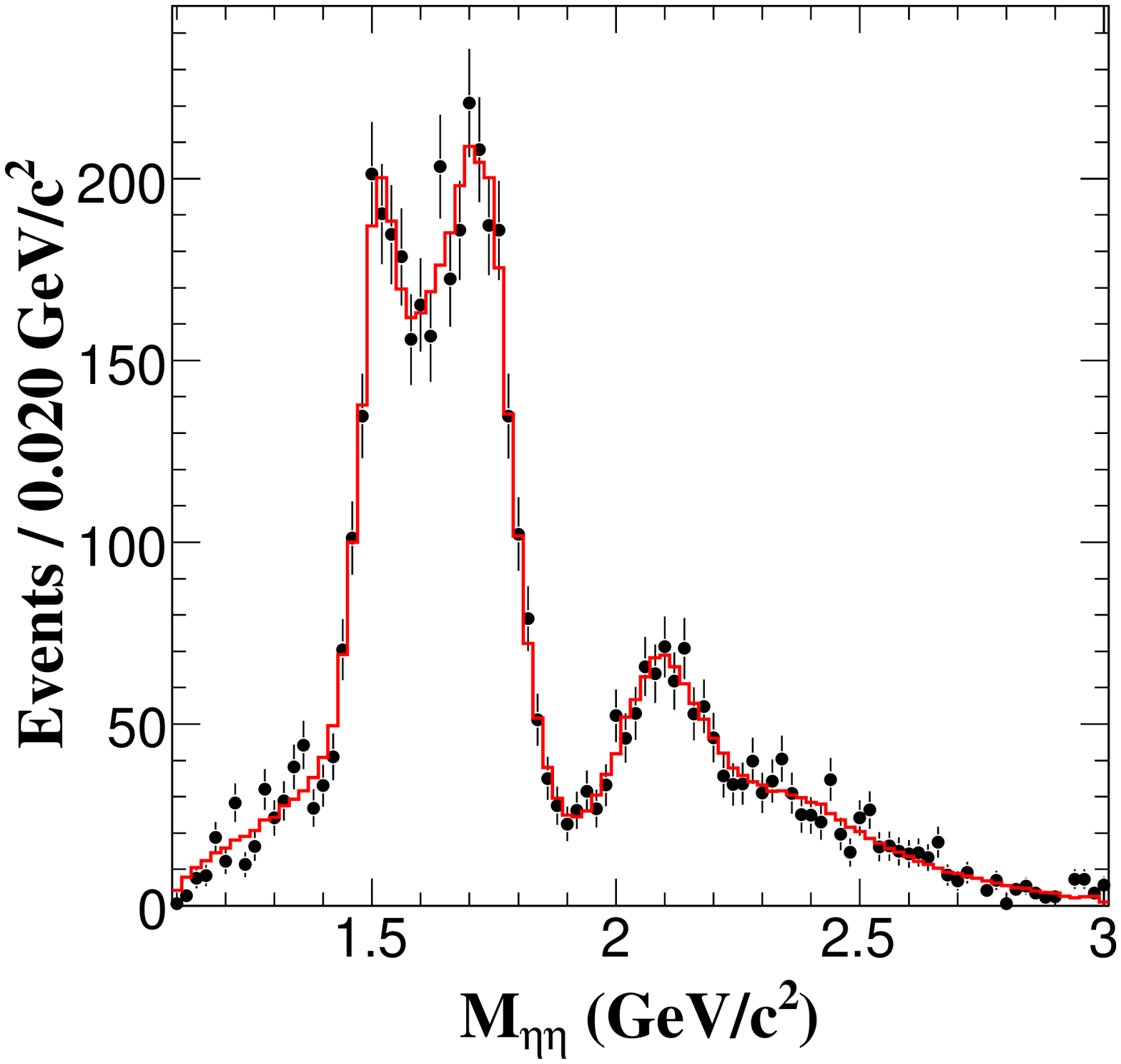}
    \put(-130,7){(a)}\put(-75,100){$\chi^{2}/N_{bin}$$=$$1.72$}}
   {\includegraphics[height=5.cm]{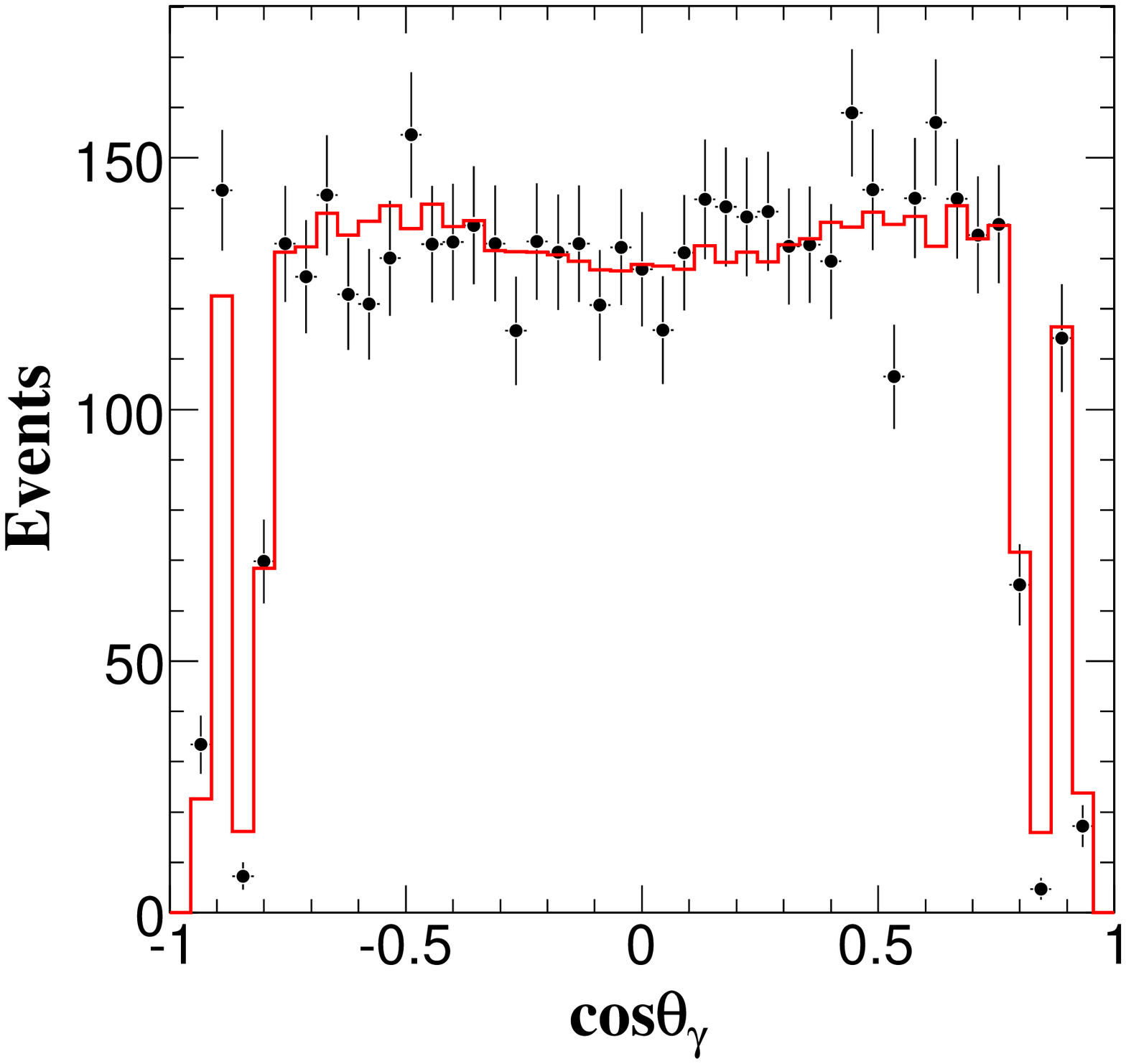}
    \put(-130,7){(b)}\put(-100,70){$\chi^{2}/N_{bin}$$=$$1.19$}}

   {\includegraphics[height=5.cm]{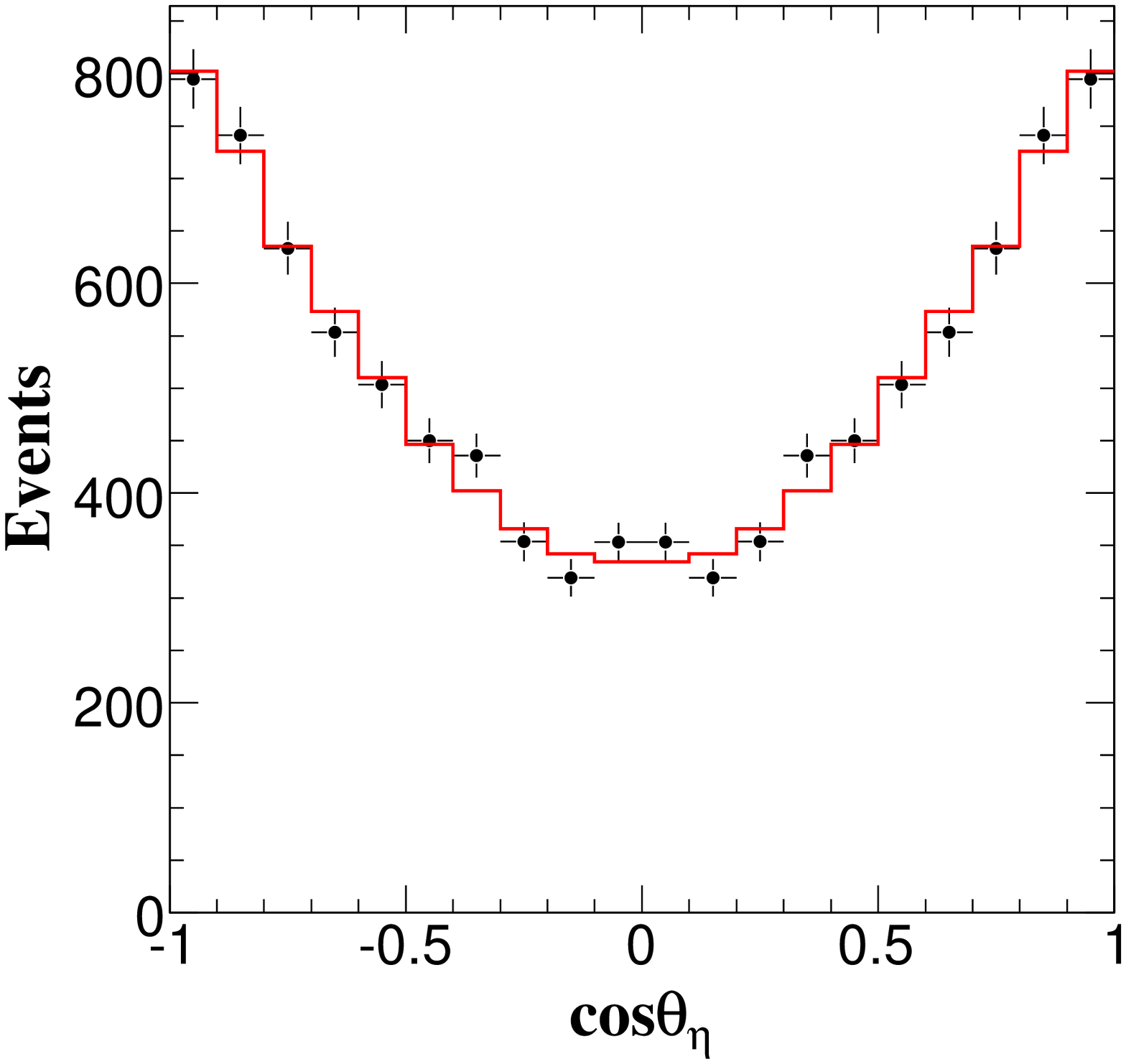}
    \put(-130,7){(c)}\put(-100,110){$\chi^{2}/N_{bin}$$=$$0.69$}}
   {\includegraphics[height=5.cm]{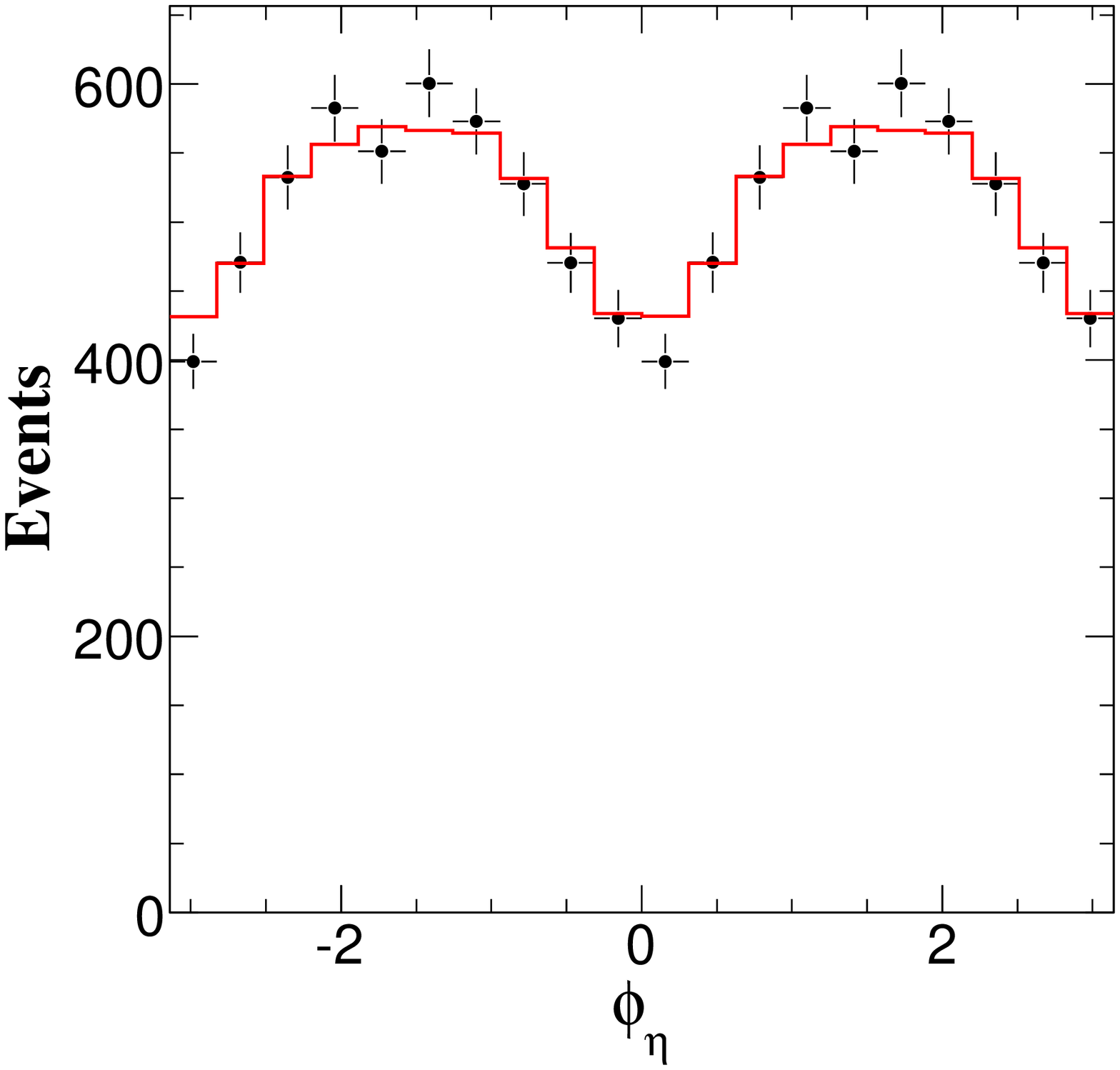}
    \put(-130,7){(d)\put(30,60){$\chi^{2}/N_{bin}$$=$$0.68$}}
   }
   \caption{Comparisons between data and PWA fit projections: (a) the
     invariant mass spectrum of $\eta\eta$, (b)-(c) the polar angle of
     the radiative photon in the $J/\psi$ rest frame and $\eta$ in the
     $\eta\eta$ helicity frame, and (d) the azimuthal angle of $\eta$ in
     the $\eta\eta$ helicity frame. The black dots with error bars are
     data with background subtracted, and the solid histograms show the
     PWA projections.}
   \vskip -0.5cm
   \label{fig:pwafitresult}
\end{figure*}

\begin{figure*}[htbp]
   \vskip -0.1cm
   \centering
   {\includegraphics[width=4.5cm,height=3.9cm]{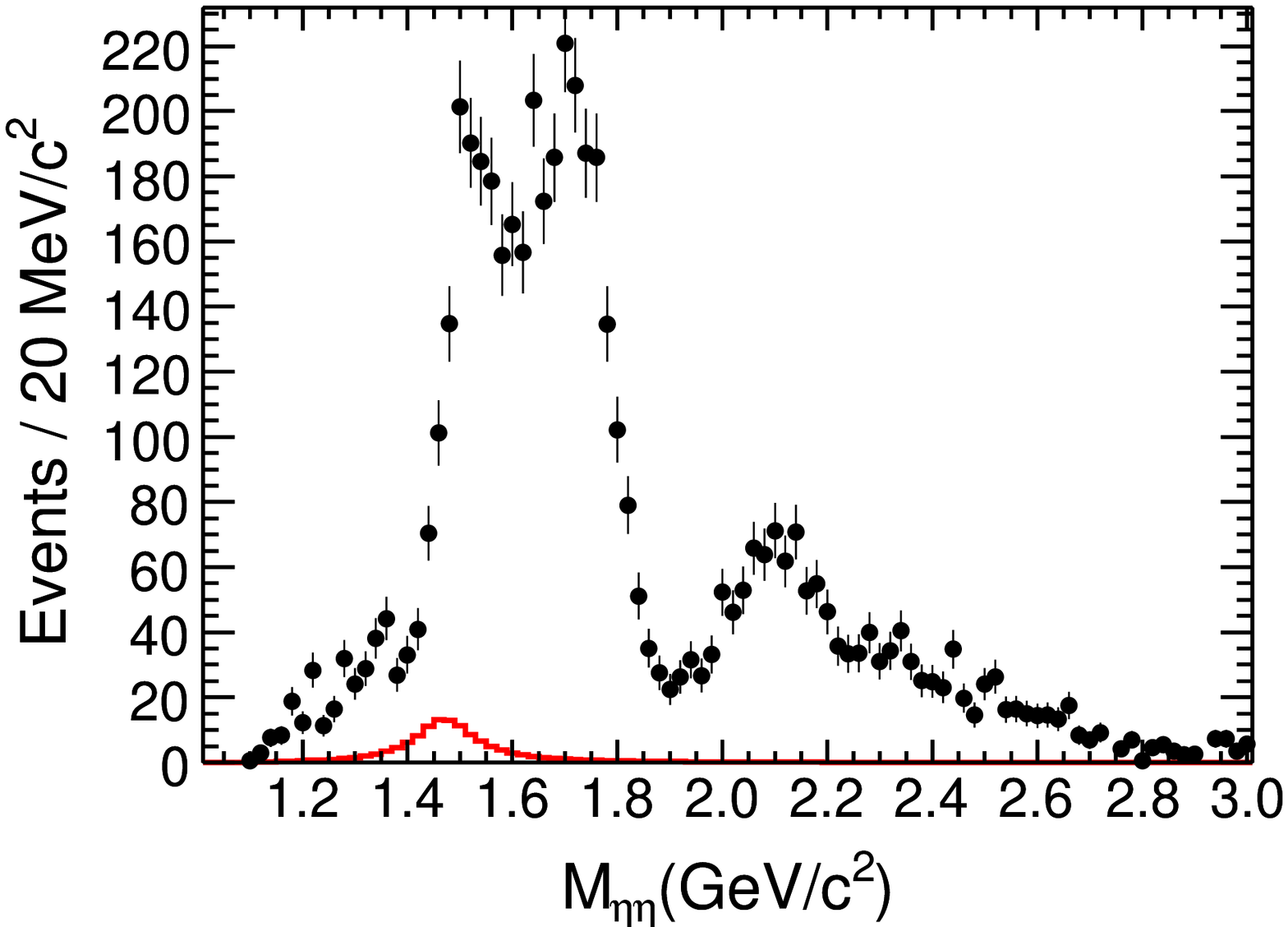}
    \put(-90,2){(a)}}
   {\includegraphics[width=4.5cm,height=3.9cm]{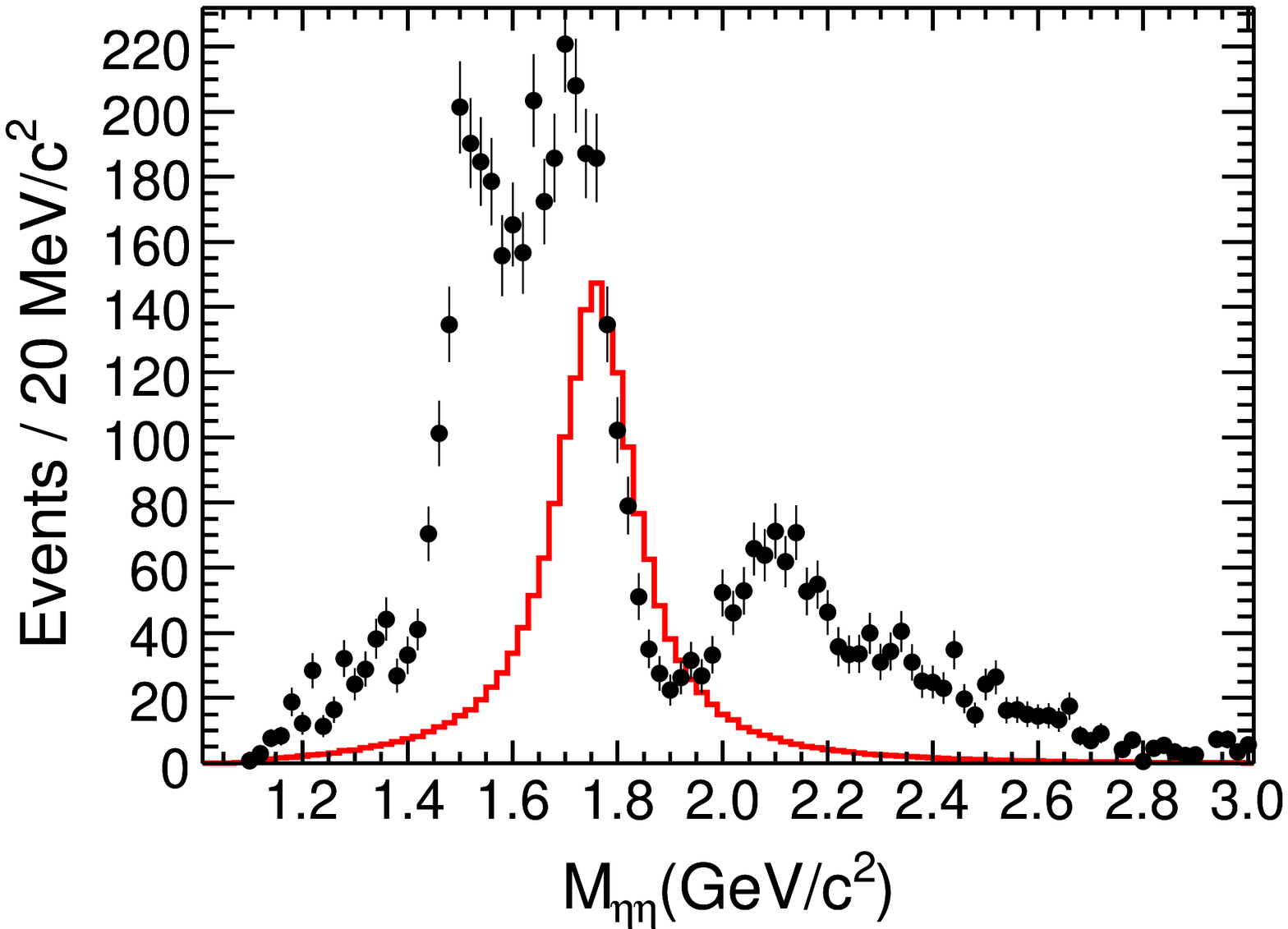}
    \put(-90,2){(b)}}
   {\includegraphics[width=4.5cm,height=3.9cm]{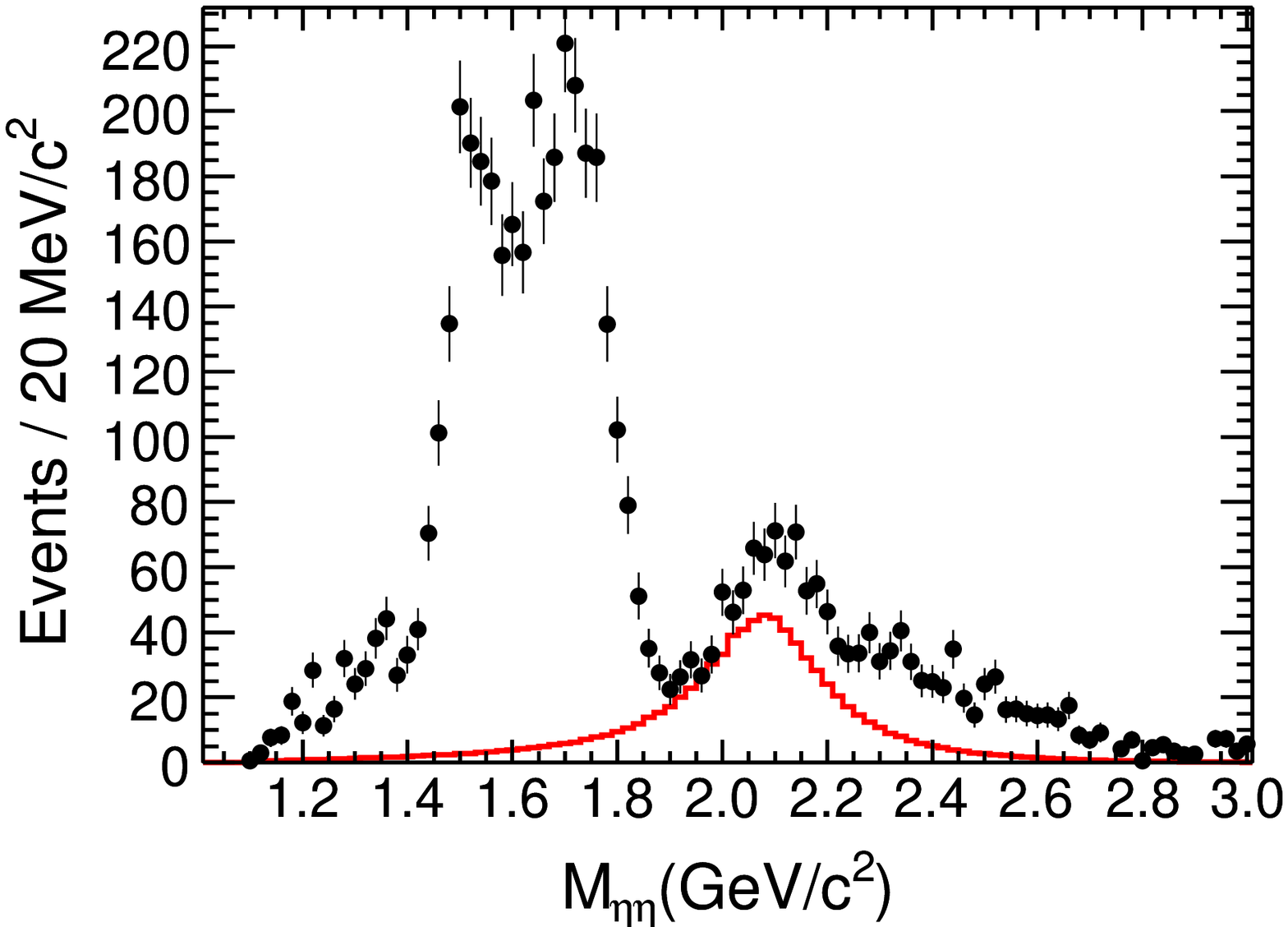}
    \put(-90,2){(c)}}
   {\includegraphics[width=4.5cm,height=3.9cm]{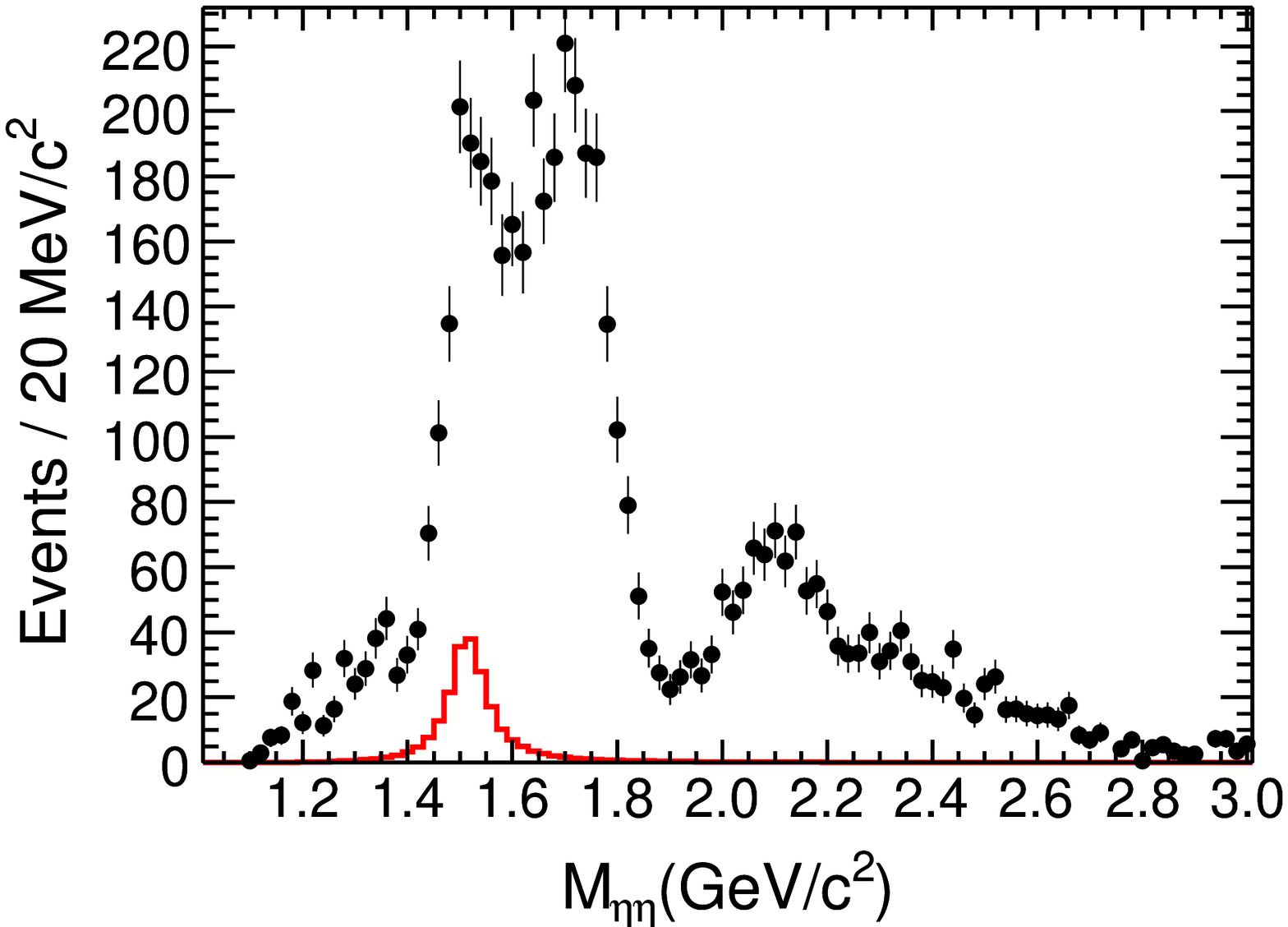}
    \put(-90,2){(d)}}
   {\includegraphics[width=4.5cm,height=3.9cm]{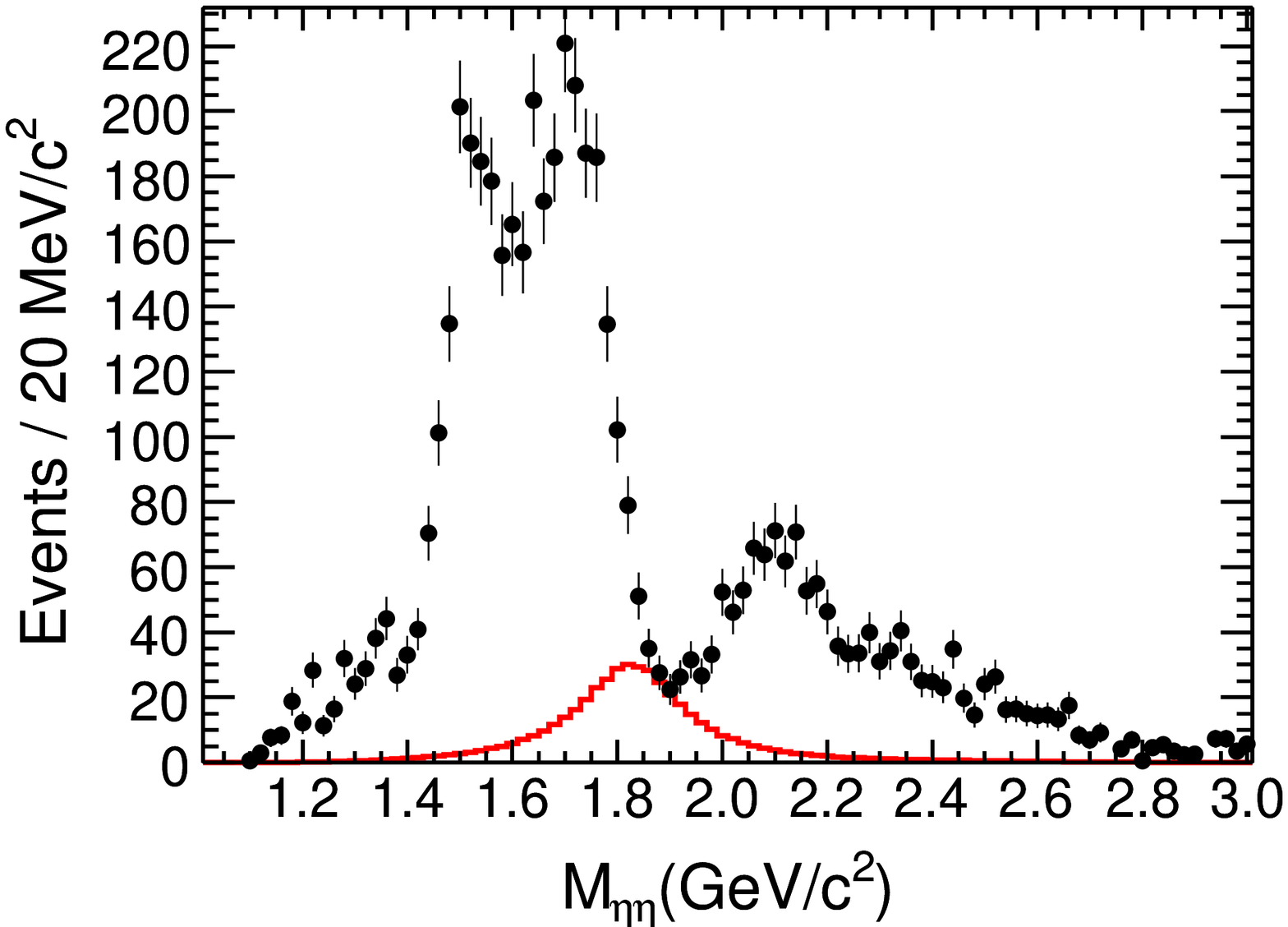}
    \put(-90,2){(e)}}
   {\includegraphics[width=4.5cm,height=3.9cm]{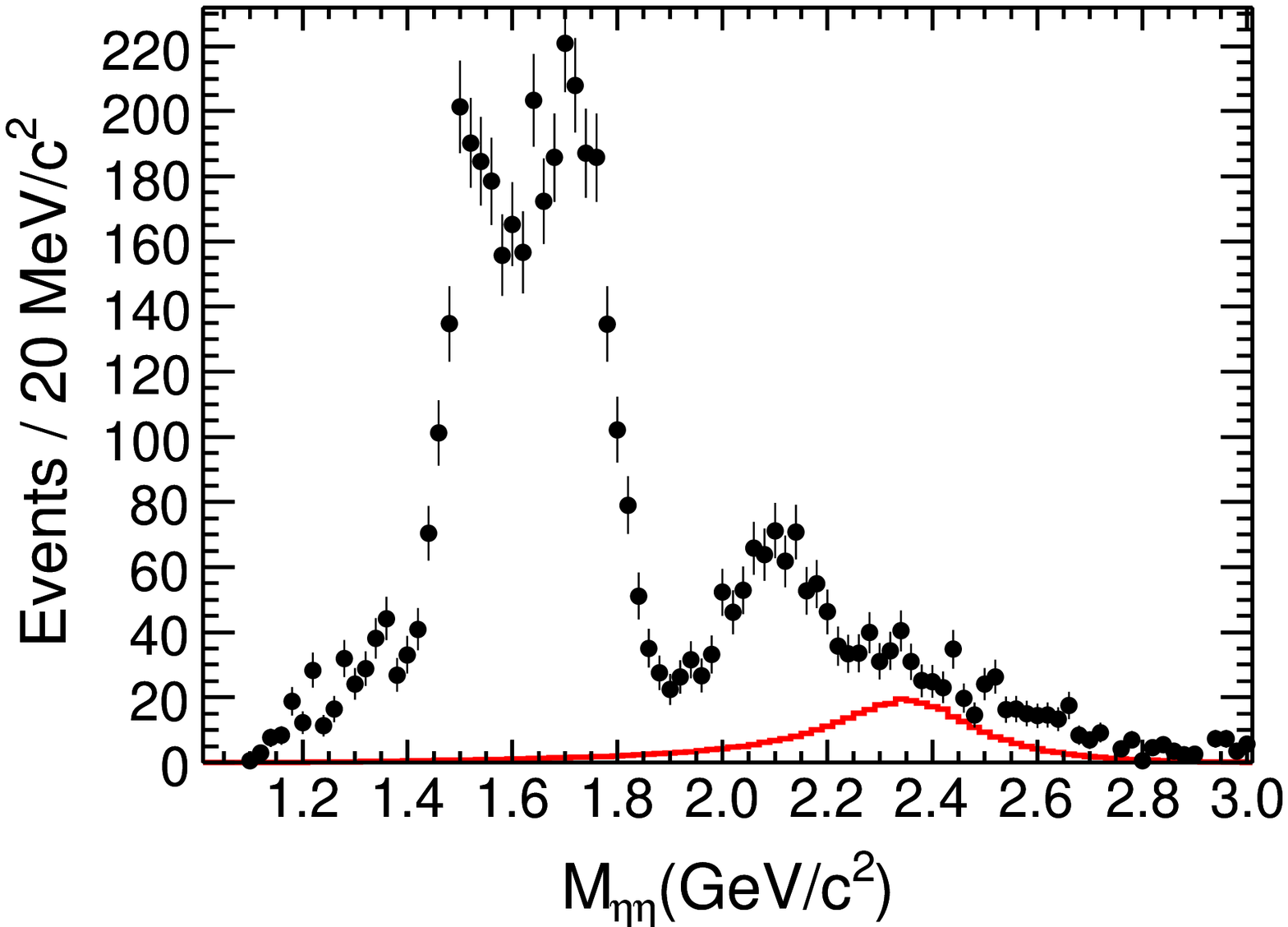}
    \put(-90,2){(f)}}
   {\includegraphics[width=4.5cm,height=3.9cm]{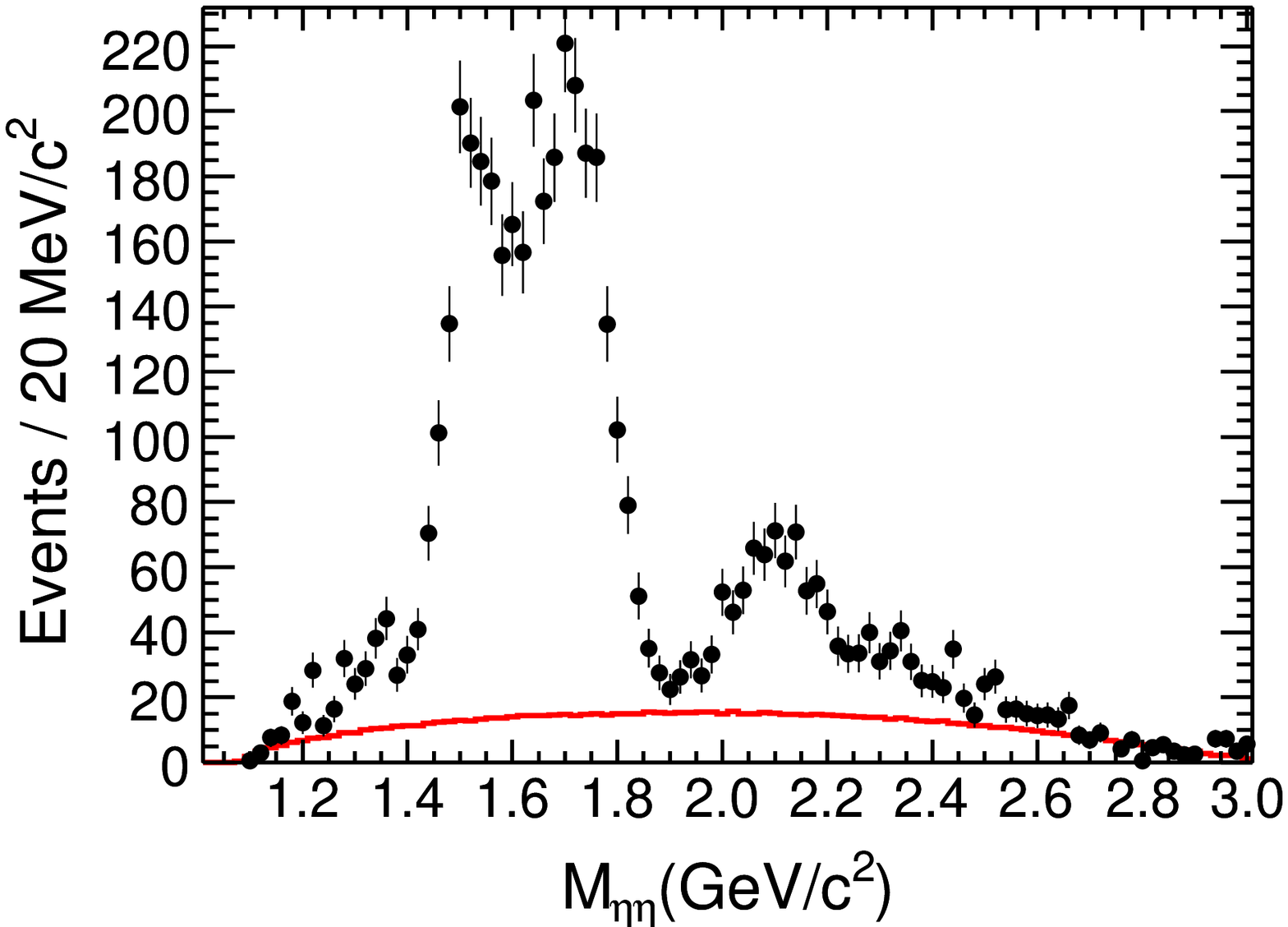}
    \put(-90,2){(g)}}
   {\includegraphics[width=4.5cm,height=3.9cm]{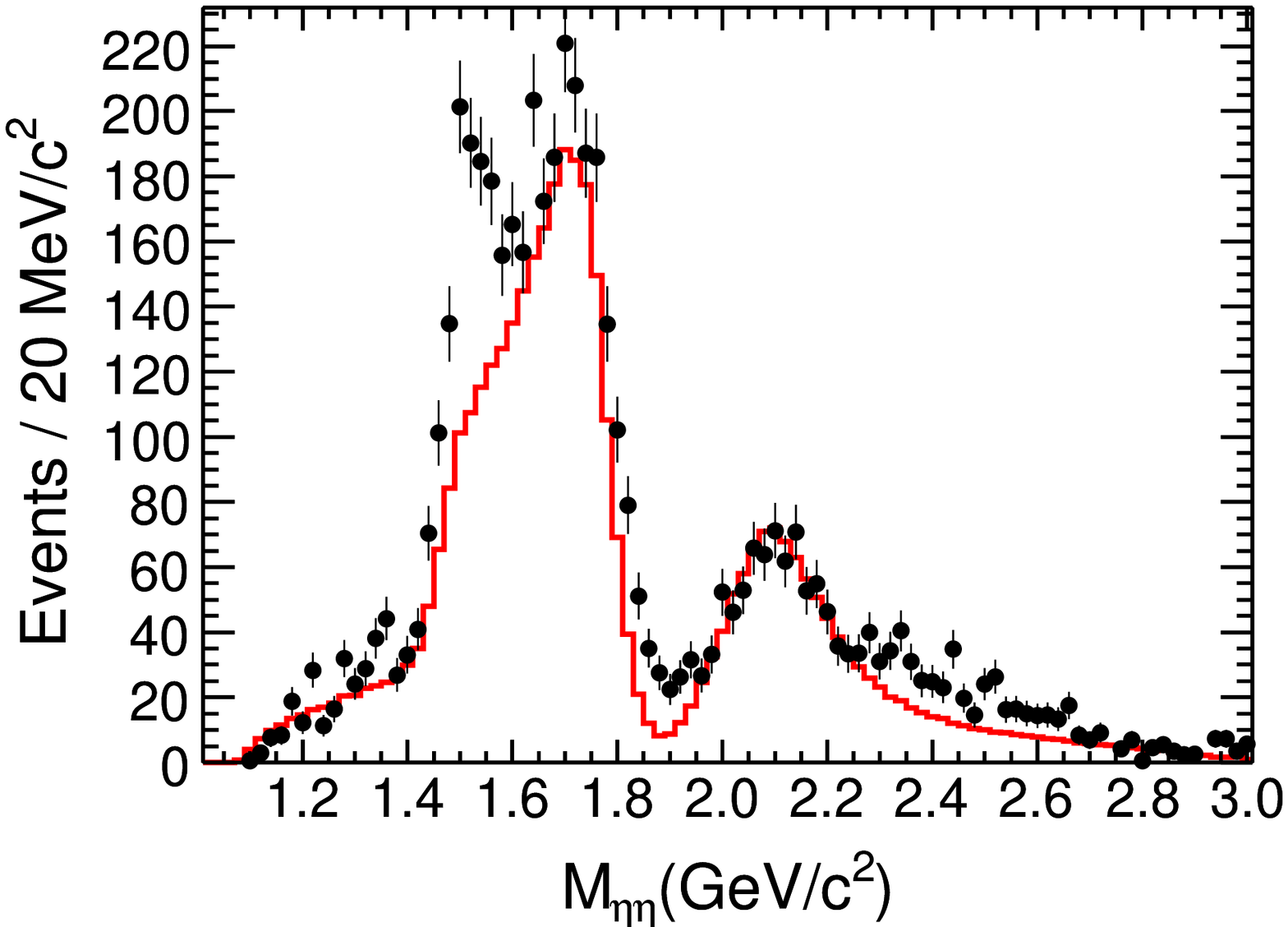}
    \put(-90,2){(h)}}
   {\includegraphics[width=4.5cm,height=3.9cm]{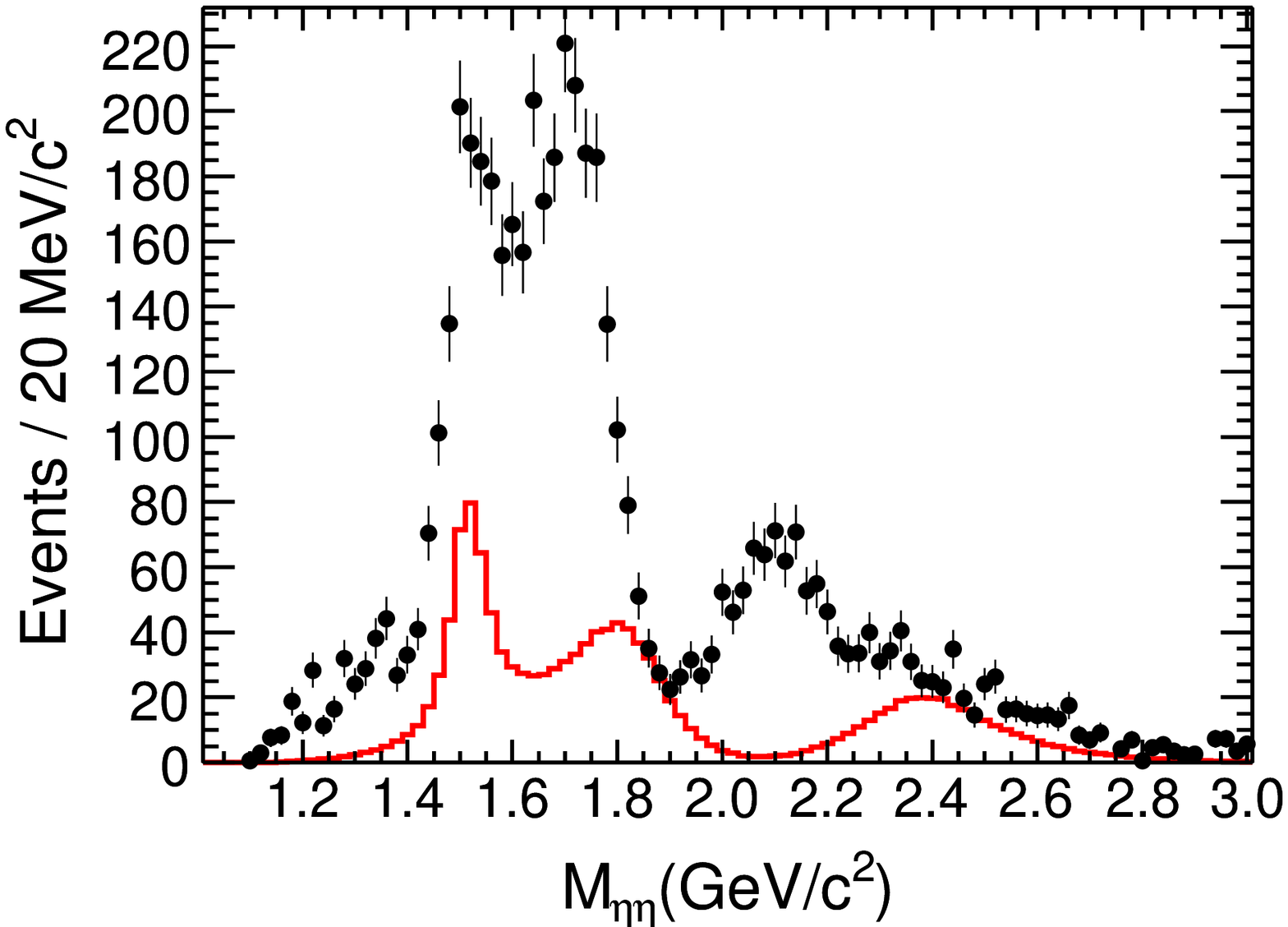}
    \put(-90,2){(i)}}

  \caption{Contribution of the components. (a) $f_{0}(1500)$, (b)
    $f_{0}(1710)$, (c) $f_{0}(2100)$, (d) $f_{2}^{'}(1525)$, (e)
    $f_{2}(1810)$, (f) $f_{2}(2340)$, (g) $0^{++}$ phase space, (h)
    total $0^{++}$ component, and (i) total $2^{++}$ component.  The dots
    with error bars are data with background subtracted, and the solid
    histograms are the projection of the PWA.}
   \vskip -0.5cm
   \label{fig:component}
\end{figure*}

The comparisons of the $\eta\eta$ invariant mass spectrum,
$\cos\theta_{\eta}$, $\cos\theta_{\gamma}$ and $\phi_{\eta}$
distributions between the data and the PWA fit projections (weighted
by MC efficiencies) are displayed in Fig.~\ref{fig:pwafitresult} (a),
(b), (c), and (d), where $\theta_{\gamma}$ is the polar angle of the
radiative photon in the $J/\psi$ rest frame, and $\theta_{\eta}$ and
$\phi_{\eta}$ are the polar angle and azimuthal angle of $\eta$ in the
$\eta\eta$ helicity frame. The PWA results provide a good description
of data. To illustrate the contributions from each component, the PWA
projections for each specific resonance are plotted (Fig.~\ref{fig:component} (a)-(f):
$f_{0}(1500)$, $f_{0}(1710)$, $f_{0}(2100)$, $f_{2}^{'}(1525)$,
$f_{2}(1810)$, $f_{2}(2340)$), $0^{++}$ phase space (Fig.~\ref{fig:component} (g)), total
$0^{++}$ component (Fig.~\ref{fig:component} (h)) and total $2^{++}$ component (Fig.~\ref{fig:component} (i)),
where the dots with error bars are data with the
background events subtracted and the solid histograms are the projections of the PWA
for the specific components.

\subsubsection{Scalar components}

The histogram in Fig.~\ref{fig:component} (h) shows the contribution
of all the scalar components, where the dominant ones are from
$f_{0}(1710)$ and $f_{0}(2100)$.  For the $f_{0}(1710)$ meson, the PWA
gives a mass of 1759$\pm6$~MeV/$c^{2}$ and a width of 172$\pm
10$~MeV/$c^{2}$ with a statistical significance of 25$\sigma$; the
mass and width are consistent with those obtained from
$J/\psi\to\gamma K\bar{K}$~\cite{bes2gammakk} and
$J/\psi\to\gamma\pi\pi$~\cite{bes2gammapipi} at BESII.  The
$f_{0}(2100)$ is observed with a statistical significance of 13.9$\sigma$,
and its mass and width are determined to be 2081$\pm
13$~MeV/$c^{2}$ and 273$^{+27}_{-24}$~MeV/$c^{2}$, respectively, which
are in agreement with previous measurements~\cite{bes1gamma4pi,Anisovich:1999fd,Anisovich:2000ut,Anisovich:2000ae}.
The product branching fractions of the $f_{0}(1710)$ and $f_{0}(2100)$
are measured to be $\br{(J/\psi\to\gamma f_{0}(1710)\to\gamma
  \eta\eta)}$ = $(2.35^{+0.13}_{-0.11})\times10^{-4}$ and
$\br{(J/\psi\to\gamma f_{0}(2100)\to\gamma
  \eta\eta)}$=$(1.13^{+0.09}_{-0.10})\times10^{-4}$, where the errors are statistical only.

The $f_{0}(1500)$ is observed with a statistical significance of 8.2$\sigma$,
but its production rate, $\br{(J/\psi\to\gamma
  f_{0}(1500)\to\gamma
  \eta\eta)}$=$(1.65^{+0.26}_{-0.31})\times10^{-5}$, is about one
order of magnitude lower than that of $f_{0}(1710)$ and $f_{0}(2100)$
since its dominant decay modes are $4\pi$ and $\pi\pi$~\cite{pdg2012}. The mass and
width obtained from the global fit are 1468$^{+14}_{-15}$~MeV/$c^{2}$
and 136$^{+41}_{-26}$~MeV/$c^{2}$, respectively, which are consistent
with the BESII measurements in
$J/\psi\to\gamma\pi\pi$~\cite{bes2gammapipi}.

The first experimental evidence for the $f_{0}(1790)$ ($M$=1790$^{+40}_{-30}$~MeV/$c^{2}$
and $\Gamma$=270$^{+60}_{-30}$~MeV/$c^{2}$) was observed in
$J/\psi\to\phi\pi\pi$~\cite{bes2phipipi}. Of interest is that no evidence
was observed in $J/\psi\to\phi K\bar{K}$~\cite{bes2phipipi}. In this
analysis, if the dominate $f_{0}(1710)$ in the basic
solution is replaced with $f_{0}(1790)$, the log likelihood is worsen by 30.
If the $f_{0}(1790)$ is included as an additional resonance in the
fit, the significance of $f_{0}(1790)$ is only 1.8$\sigma$, which
indicates that the not coupled $f_{0}(1790)$ is either suppressed in radiative decays
or not coupled strongly to $\eta\eta$.

To evaluate the contributions from other scalar mesons, $f_{0}(1370)$,
$f_{0}(2020)$, $f_{0}(2200)$ and $f_{0}(2330)$, the PWA was performed
including them, and none of them has significance greater than 5.0$\sigma$.
Therefore, they are not included in the basic solution.

\subsubsection{Tensor components}

The total contribution from the tensor components is shown as the
histogram in Fig.~\ref{fig:component} (i), where the peak around 1.5~GeV/$c^{2}$
is dominated by the well-established resonance
$f_{2}^{'}(1525)$ and the tensor components contributing to the bump
around 2.1~GeV/$c^{2}$ are from $f_2(1810)$ and $f_{2}(2340)$. The
fitted mass and width of $f_{2}^{'}(1525)$ are
1513$\pm5$~MeV/$c^{2}$ and 75$^{+12}_{-10}$~MeV/$c^{2}$,
respectively, which are consistent with the world average
values~\cite{pdg2012}, and the product branching fraction is
calculated to be $\br{(J/\psi\to\gamma
  f_{2}^{'}(1525)\to\gamma\eta\eta)}$=$(3.42^{+0.43}_{-0.51})\times10^{-5}$. If
$f_{2}^{'}(1525)$ is replaced with another tensor meson close to 1.5~GeV/$c^{2}$,
$f_2(1565)$, the log likelihood is worse by 18. The PWA
is also performed including $f_2(1565)$ as an additional resonance,
and its statistical significance is only 2.0$\sigma$.

The global fit shows that there is a tensor component around
1.8~GeV/$c^{2}$ with a statistical significance of 6.4$\sigma$, and
its mass and width are determined to be 1822$^{+29}_{-24}$~MeV/$c^{2}$
and 229$^{+52}_{-42}$~MeV/$c^{2}$, respectively, which is likely to be
the $f_{2}(1810)$. However the changes of the log likelihood value are
only 0.8 or 0.7, if we replace it with the $f_{2}(1910)$ or
$f_{2}(1950)$, respectively, using the world average values for their
masses and widths~\cite{pdg2012}, which indicates that we cannot
distinguish it from $f_{2}(1810)$, $f_{2}(1910)$ and $f_{2}(1950)$
with the present statistics. In this analysis, this tensor component
is denoted as $f_2(1810)$, and the ambiguous assignment of
$f_{2}(1810)$ or $f_{2}(1950)$ is considered as a source of systematic
error.

To investigate contributions from other possible tensor resonances,
$f_{2}(2010)$, $f_{2}(2150)$, $f_{J}(2220)$, $f_{2}(2300)$ and
$f_{2}(2340)$, the fits were performed with alternative combinations,
and the statistical significances of $f_{2}(2010)$, $f_{2}(2150)$ and
$f_{J}(2220)$ are all less than 5.0$\sigma$, and the best fit favors
the presence of $f_{2}(2340)$ (the statistical significance is 7.6$\sigma$)
with a mass of 2362$^{+31}_{-30}$~MeV/$c^{2}$, a width of
334$^{+62}_{-54}$~MeV/$c^{2}$, and a product branching fraction of
$\br{(J/\psi\to\gamma f_{2}(2340)\to\gamma
  \eta\eta)}$=$(5.60^{+0.62}_{-0.65})\times10^{-5}$. Since the mass of
$f_{2}(2300)$ is close to $f_{2}(2340)$, an attempt was made to replace
$f_{2}(2340)$ with $f_{2}(2300)$ by fixing its mass and width to those
in PDG~\cite{pdg2012}, and the log likelihood value is worse by
15. The narrow $f_{J}(2220)$ (also known as $\xi(2230)$), which
was reported by MarkIII~\cite{Baltrusaitis:1985pu} and
BES~\cite{Bai:1998tx}, is also studied. In this analysis no evident narrow peak around
2.2~GeV/$c^{2}$ over the broad bump is observed in the $\eta\eta$ mass
spectrum shown in Fig.~\ref{fig:data}~(c). When the $f_{J}(2220)$
is included in the PWA, the statistical significance is found to be 0.4$\sigma$.

\subsubsection{Non-resonant contribution and $J/\psi\to\phi\eta$}
In the analysis, the non-resonant contribution in the decay
$J/\psi\rightarrow\gamma\eta\eta$ is described with $0^{++}$ phase
space, with a statistical significance of 12.4$\sigma$, and the product branching fraction is
calculated to be $(1.47^{+0.01}_{-0.02})\times10^{-4}$.
An alternative fit is made by replacing the $0^{++}$ phase
space with $2^{++}$ phase space, and the log likelihood value is worsen
by 30. In addition to $0^{++}$ phase space, a $2^{++}$ phase space
component to describe the non-resonant contribution was used, but the
significance of the additional $2^{++}$ phase space is less than 4.0$\sigma$.
The impact from the uncertainty of the non-resonant
contribution is taken as a source of systematic error.

\section{Systematic error}
The systematic sources and their corresponding contributions to the
measurements of mass, width and branching fractions are described below.

\begin{itemize}
\item Background uncertainty.  The background events
estimated with the $\eta$ mass sidebands are included in the global
fit with negative weights.  To estimate the systematic error, the
global fit was done with background events from different $\eta$ mass
sideband regions, and the changes of results are assigned as the
systematic errors.

\item Uncertainty from extra components.  As mentioned above, possible
extra $0^{++}$, $2^{++}$ and $4^{++}$ components with low significance were
removed from the global fit. The changes of results caused by including
them in the global fit are assigned as the systematic errors.

\item Uncertainty from resonance parameters. To estimate the impact of
one specific resonance on the others, the optimized mass and width of
each resonance were varied by one standard deviation (statistical
error only), and then the global fit was redone. The differences
between the results with and without the variation of the resonance
parameters are assigned as the systematic errors. As discussed above,
the $f_{2}(1810)$ cannot be distinguished from $f_{2}(1910)$ or
$f_{2}(1950)$. Therefore the fits were redone fixing the mass and
width to be the values of $f_{2}(1810)$ and $f_{2}(1950)$ in
PDG~\cite{pdg2012}, respectively, and the maximum changes of the
results are regarded as the systematic uncertainties.

\item To estimate the uncertainty
 from $J/\psi\to\phi\eta$, an alternative fit was performed without the contribution of
$J/\psi\to\phi\eta$, and the changes of results are taken as the
systematic errors.

\item Mass resolution. In the global fit, the mass resolution is not
considered to simplify the analysis. In order to estimate its possible
impact on the fitted resonance parameters, a test was made by smearing
the line shape of each resonance from the global fit with the
corresponding mass resolution obtained from MC simulation. The impact
on a resonance with a width greater than 100 MeV/$c^2$ is less than
2\%, which is negligible compared with uncertainties from other
sources. For $f^{'}_{2}(1525)$, the width is smeared by 5 MeV/$c^2$
with respect to the PWA result
75$^{+12}_{-10}$~(stat.)~MeV/$c^{2}$. The difference is considered as
a source of systematic error to the width measurement.

\item Phase space description. The uncertainty from the description of
the non-resonant contribution is estimated from an alternative fit by
including both $0^{++}$ and $2^{++}$ phase space.

\item Breit-Wigner formula. The changes of the fit
results caused by replacing the constant width Breit-Wigner with a
kinematic dependent width Breit-Wigner~\cite{depbw} are taken as the
uncertainties from different resonance parameterizations.

\end{itemize}
In addition to the above systematic sources, the systematic errors
from the event selection criteria, trigger efficiency and the number
of $J/\psi$ events, which are summarized in Table~\ref{err}, are also
included in the branching fraction measurements.

%%%%%%%%%%%%%%%%%%%%%%%%%%%%%%%%%%%%%%%%%%%%%%%%%%%%%%%%%%%%%%%%%%%%%%%%%%%%%%%%%%%%%%%%%%%%%%%%%%%%%%%%%%%%%%%%%%%%%%%%%%%%%%%%%%%%%%%%%%%
%% Table II
%%%%%%%%%%%%%%%%%%%%%%%%%%%%%%%%%%%%%%%%%%%%%%%%%%%%%%%%%%%%%%%%%%%%%%%%%%%%%%%%%%%%%%%%%%%%%%%%%%%%%%%%%%%%%%%%%%%%%%%%%%%%%%%%%%%%%%%%%%%
\begin {table*}[htb]
\begin {center}
\caption {Summary of the systematic errors from the event selection.}
\begin {tabular}{ccc|rrr}
\hline\hline Error sources  &Systematic error(\%) \\\hline

\hline Photon efficiency  &5.0\\%\hline

Kinematic fit  &6.5\\%\hline

$\eta$ selection &0.8\\%\hline

Number of $J/\psi$ events &1.24\\\hline

Total  &8.3\\\hline\hline

\end {tabular}
\label{err}
\end {center}
\end {table*}

\begin{itemize}

\item Photon detection. For the decay mode analyzed in this paper,
five photons are involved in the final states.
The uncertainty due to photon detection and photon conversion is 1\% per
photon. This is determined from studies of photon detection
efficiencies in well understood decays such as $J/\psi\to\rho^0\pi^0$ and
study of photon conversion via $e^+e^-\to\gamma\gamma$~\cite{Ablikim:2011kv,Ablikim:2010zn}.

\item Trigger efficiency. The trigger efficiency of the BESIII
detector was found to be very close to 100\% from studies using
different samples selected from $J/\psi$ or $\psi(2S)$ decays.
Therefore, the trigger efficiency is assumed to be 100\%
in the calculation of the branching fractions, and the systematic
error from this source is neglected.

\item Kinematic fit and $\eta$ selection. The systematic error from
the kinematic fit is studied with the clean channel
$\psi'\to\gamma\chi_{c0}$ ($\chi_{c0}\to\eta\eta$), as described in
Ref.~\cite{Ablikim:2010zn}. The efficiency is defined as the ratio of
$\chi_{c0}$ yield with and without the kinematic requirement of
$\chi^2_{4C}<50 $, where the $\chi_{c0}$ yield is obtained by fitting
the $\eta\eta$ mass spectrum with the MC signal shape and a
second-order polynomial. The difference between data and MC
simulation, 6.5\%, is taken to be the systematic error. Similarly the
systematic error from $\eta$ selection criteria is estimated to be 0.8\%.

\item Number of $J/\psi$ events. In the calculation of branching
fractions, the number of $J/\psi$ events,
$(225.3\pm2.8)\times10^{6}$~\cite{jpsinumber}, determined from
$J/\psi$ inclusive hadronic decays, was used, and its uncertainty,
1.24\%, is taken as the systematic error.
\end{itemize}

%%%%%%%%%%%%%%%%%%%%%%%%%%%%%%%%%%%%%%%%%%%%%%%%%%%%%%%%%%%%%%%%%%%%%%%%%%%%%%%%%%%%%%%%%%%%%%%%%%%%%%%%%%%%%%%%%%%%%%%%%%%%%%%%%%%%%%%%%%%
%% Table III
%%%%%%%%%%%%%%%%%%%%%%%%%%%%%%%%%%%%%%%%%%%%%%%%%%%%%%%%%%%%%%%%%%%%%%%%%%%%%%%%%%%%%%%%%%%%%%%%%%%%%%%%%%%%%%%%%%%%%%%%%%%%%%%%%%%%%%%%%%%

\begin{table*}
\caption{Summary of the systematic error sources and their corresponding contributions to
masses and widths of the resonances X (MeV/$c^2$), which are denoted as
$\Delta M$ and $\Delta \Gamma$, respectively.}
\begin{ruledtabular}
{\scriptsize
\begin{tabular}{lcccccccccccc}
Systematic error& \multicolumn{2}{c}{$f_{0}(1500)$}&
\multicolumn{2}{c}{$f_{2}^{'}(1525)$}& \multicolumn{2}{c}{$f_{0}(1710)$}&
\multicolumn{2}{c}{$f_{2}(1810)$}& \multicolumn{2}{c}{$f_{0}(2100)$}&
\multicolumn{2}{c}{$f_{2}(2340)$} \\
&$\Delta M$ & $\Delta \Gamma$&
$\Delta M$  & $\Delta \Gamma$&
$\Delta M$  & $\Delta \Gamma$&
$\Delta M$  & $\Delta \Gamma$&
$\Delta M$  & $\Delta \Gamma$&
$\Delta M$  & $\Delta \Gamma$\\
\hline
Background uncertainty    &$^{+18}_{-8}$  &-46 &+1  &$^{+6}_{-4}$ &$^{+11}_{-13}$  &$^{+19}_{-8}$ &$^{+55}_{-39}$ &$^{+61}_{-4}$ &$^{+19}_{-32}$ &+38  &+93 &$^{+43}_{-41}$\\
Extra resonances  &$^{+9}_{-73}$ &$^{+7}_{-84}$ &$^{+2}_{-9}$ &$^{+10}_{-4}$ &$^{+8}_{-18}$  &$^{+19}_{-11}$ &$^{+19}_{-36}$  &$^{+16}_{-141}$  &$^{+7}_{-9}$  &$^{+50}_{-4}$ &$^{+76}_{-9}$  &$^{+7}_{-2}$\\

Resonance parameters &$^{+11}_{-12}$  &$^{+27}_{-25}$ &$^{+3}_{-4}$ &$^{+8}_{-5}$   &$^{+4}_{-11}$ &$^{+17}_{-7}$  &$^{+24}_{-18}$  &$^{+61}_{-26}$  &$^{+13}_{-12}$ &$^{+26}_{-23}$ &$^{+55}_{-62}$ &$^{+157}_{-87}$\\

$J/\psi \to \phi \eta$   &-1 &-10 &0  &+2 &0  &+3 &+11  &+3 &-4 &+12 &+16 &+26\\
Phase space description  &0 &+3 &-1  &-1 &-1  &0 &-11  &+7 &-3 &+10 &+38 &-21\\
Breit-Wigner formula    &0 &-6 &+1  &+6 &-5 &-5 &+15 &-58  &+2 &+4 &+20 &-18 \\\hline

Total &$^{+23}_{-74}$ &$^{+28}_{-100}$ &$^{+4}_{-10}$ &$^{+16}_{-8}$ &$^{+14}_{-25}$ &$^{+32}_{-16}$ &$^{+66}_{-57}$ &$^{+88}_{-155}$ &$^{+24}_{-36}$ &$^{+70}_{-23}$ &$^{+140}_{-63}$ &$^{+165}_{-100}$\\

\end{tabular}
}
\end{ruledtabular}
\label{errpwa1}
\end{table*}

%%%%%%%%%%%%%%%%%%%%%%%%%%%%%%%%%%%%%%%%%%%%%%%%%%%%%%%%%%%%%%%%%%%%%%%%%%%%%%%%%%%%%%%%%%%%%%%%%%%%%%%%%%%%%%%%%%%%%%%%%%%%%%%%%%%%%%%%%%%
%% Table IV
%%%%%%%%%%%%%%%%%%%%%%%%%%%%%%%%%%%%%%%%%%%%%%%%%%%%%%%%%%%%%%%%%%%%%%%%%%%%%%%%%%%%%%%%%%%%%%%%%%%%%%%%%%%%%%%%%%%%%%%%%%%%%%%%%%%%%%%%%%%

\begin {table*}[htb]
\begin {center}
\caption{Summary of the systematic error sources and their corresponding contributions to
the branching fractions of $J/\psi\to\gamma$X, $X\to\eta\eta$~(\%), which are denoted as
$\Delta$$\br$.}
%{\scriptsize
\begin{tabular}{lcccccc}
\hline\hline Systematic error  &$\Delta$$\br$($f_{0}(1500)$)  &$\Delta$$\br$($f_{2}^{'}(1525)$)  &$\Delta$$\br$($f_{0}(1710)$)  &$\Delta$$\br$($f_{2}(1810)$)  &$\Delta$$\br$($f_{0}(2100)$)  &$\Delta$$\br$($f_{2}(2340)$) \\
\hline Event selection &$\pm8.3$ &$\pm8.3$ &$\pm8.3$ &$\pm8.3$ &$\pm8.3$ &$\pm8.3$ \\

Background uncertainty    &$^{+20.6}_{-46.1}$  &$^{+23.5}_{-34.8}$  &$^{+35.4}_{-15.1}$  &$^{+5.1}_{-34.5}$   &$^{+46.8}_{-9.7}$   &$^{+24.9}_{-0.9}$\\
Extra resonances          &$^{+11.1}_{-56.3}$  &+21.9               &$^{+33.2}_{-23.9}$  &$^{+20.3}_{-19.0}$  &$^{+26.9}_{-16.2}$  &$^{+6.0}_{-27.8}$\\
Resonance parameters      &$^{+18.0}_{-41.6}$  &$^{+21.6}_{-12.0}$  &$^{+17.0}_{-8.0} $  &$^{+58.2}_{-14.4}$  &$^{+11.3}_{-11.1}$  &$^{+24.3}_{-19.3}$\\
$J/\psi \to \phi \eta$    &-7.1  &+0.6  &+7.6  &-6.4  &+8.7  &+5.0  \\
Phase space description   &-1.6  &-3.2  &-0.5  &+10.7 &-1.0  &+21.1 \\
Breit-Wigner formula      &-6.3  &+6.8  &-8.4  &-4.9  &-7.4  &-12.5 \\\hline

Total   &$^{+30.7}_{-84.8}$  &$^{+40.2}_{-37.9}$  &$^{+52.6}_{-31.7}$  &$^{+63.3}_{-43.5}$  &$^{+56.5}_{-24.6}$  &$^{+42.3}_{-37.0}$  \\\hline\hline

\end{tabular}
%}
\label{errpwa2}
\end {center}
\end{table*}

The systematic error sources and their contributions studied above are
all summarized in Table~\ref{errpwa1} and Table~\ref{errpwa2}, in which the systematic error
from event selection includes the contributions from photon detection
efficiency, kinematic fit, $\eta$ selection and the number of $J/\psi$
events listed in Table~\ref{err}.  The total systematic error is the
sum of them added in quadrature.

%%%%%%%%%%%%%%%%%%%%%%%%%%%%%%%%%%%%%%%%%%%%%%%%%%%%%%%%%%%%%%%%%%%%%%%%%%%%%%%%%%%%%%%%%%%%%%%%%%%%
\section{Summary}
Using 225 million $J/\psi$ events collected with the BESIII detector,
a PWA of $J/\psi\to\gamma\eta\eta$ has been performed, and the results
are summarized in Table~\ref{mwb}. The scalar contributions are mainly
from $f_{0}(1500)$, $f_{0}(1710)$ and $f_{0}(2100)$, while no evident
contributions from $f_{0}(1370)$ and $f_0(1790)$ are seen.  Recently,
the production rate of the pure gauge scalar glueball in $J/\psi$
radiative decays predicted by the lattice QCD~\cite{Gui:2012gx} was
found to be compatible with the production rate of $J/\psi$ radiative
decays to $f_{0}(1710)$; this suggests that $f_{0}(1710)$ has a larger
overlap with the glueball compared to other glueball candidates
(eg. $f_{0}(1500)$). In this analysis, the production rate of
$f_{0}(1710)$ and $f_{0}(2100)$ are both about one order of magnitude
larger than that of the $f_{0}(1500)$ and no clear evidence is found
for $f_{0}(1370)$, which are both consistent with, at least not
contrary to, lattice QCD predictions.

Studies using data from $\bar{p}p$
annihilation~\cite{Anisovich:2000ut,Anisovich:2000ae} show that the
$f_{0}(2100)$ has strong coupling to $\eta\eta$, but much weaker to
$\pi\pi$, which indicates an exotic $f_{0}(2100)$ decay
pattern. Searching for more decay modes of $f_{0}(2100)$ in $J/\psi$
radiative decays may help to clarify its nature.

The tensor components, which are dominantly from $f_{2}^{'}(1525)$,
$f_{2}(1810)$ and $f_2(2340)$, also have a large contribution in
$J/\psi\to\gamma\eta\eta$ decays. The significant contribution
from $f_{2}^{'}(1525)$ is shown as a clear peak in the $\eta\eta$ mass
spectrum; a tensor component exists in the mass region from 1.8~GeV/$c^2$
to 2~GeV/$c^2$, although we cannot distinguish $f_{2}(1810)$ from
$f_{2}(1910)$ or $f_{2}(1950)$; and the PWA requires a
strong contribution from $f_2(2340)$, although the possibility of
$f_2(2300)$ cannot be ruled out. For the narrow $f_J(2220)$, no
evident peak is observed in the $\eta\eta$ mass spectrum. We have
also tried to add it in the analysis, but its statistical
significance is quite small, just 0.4$\sigma$.

%%%%%%%%%%%%%%%%%%%%%%%%%%%%%%%%%%%%%%%%%%%%%%%%%%%%%%%%%%%%%%%%%%%%%%%%%%%%%%%%%%%%%%%%%%%%%%%%%%%%
\section{Acknowledgments}

The BESIII collaboration thanks the staff of BEPCII and the computing center for their hard efforts. This work is supported in part by the Ministry of Science and Technology of China under Contract No. 2009CB825200; National Natural Science Foundation of China (NSFC) under Contracts Nos. 10625524, 10821063, 10825524, 10835001, 10935007, 11125525; Joint Funds of the National Natural Science Foundation of China under Contracts Nos. 11079008, 11179007; the Chinese Academy of Sciences (CAS) Large-Scale Scientific Facility Program; CAS under Contracts Nos. KJCX2-YW-N29, KJCX2-YW-N45; 100 Talents Program of CAS; German Research Foundation DFG under Contract No. Collaborative Research Center CRC-1044; Istituto Nazionale di Fisica Nucleare, Italy; Ministry of Development of Turkey under Contract No. DPT2006K-120470; U. S. Department of Energy under Contracts Nos. DE-FG02-04ER41291, DE-FG02-94ER40823; U.S. National Science Foundation; University of Groningen (RuG) and the Helmholtzzentrum fuer Schwerionenforschung GmbH (GSI), Darmstadt; WCU Program of National Research Foundation of Korea under Contract No. R32-2008-000-10155-0

%\end{newpage}
%\begin{newpage}


\begin{thebibliography}{99}

\bibitem{Close:1987er}
F.~E.~Close, Rept.\ Prog.\ Phys.\  {\bf 51}, 833 (1988).

\bibitem{Godfrey:1998pd}
S.~Godfrey and J.~Napolitano,
%``Light meson spectroscopy,''
Rev.\ Mod.\ Phys.\  {\bf 71}, 1411 (1999).

\bibitem{Amsler:2004ps}
C.~Amsler and N.~A.~Tornqvist,
%``Mesons beyond the naive quark model,''
Phys.\ Rept.\  {\bf 389}, 61 (2004).

\bibitem{Klempt:2007cp}
E.~Klempt and A.~Zaitsev,
%``Glueballs, Hybrids, Multiquarks. Experimental facts versus QCD inspired concepts,''
Phys.\ Rept.\  {\bf 454}, 1 (2007).

\bibitem{Crede:2008vw}
V.~Crede and C.~A.~Meyer,
%``The Experimental Status of Glueballs,''
Prog.\ Part.\ Nucl.\ Phys.\  {\bf 63}, 74 (2009).

\bibitem{Chen:2005mg}
Y.~Chen {\it et al.},
%``Glueball spectrum and matrix elements on anisotropic lattices,''
Phys.\ Rev.\ D {\bf 73}, 014516 (2006).

\bibitem{Gregory:2012hu}
E.~Gregory {\it et al.},
%``Towards the glueball spectrum from unquenched lattice QCD,''
JHEP {\bf 1210}, 170 (2012).

\bibitem{Edwards:1981ex}
C.~Edwards {\it et al.},
%``Observation of an eta eta Resonance in J/psi Radiative Decays,''
Phys.\ Rev.\ Lett.\  {\bf 48}, 458 (1982).

%\cite{Ablikim:2012cn}
\bibitem{jpsinumber}
M.~Ablikim {\it et al.}  (BESIII Collaboration),
%``Determination of the number of $J/\psi$ events with $J/\psi \rightarrow \, inclusive$ decays,''
Chin.\ Phys.\ C {\bf 36}, 915 (2012).

\bibitem{bepc2}
J.~Z. Bai {\it et al.} (BES Collaboration), Nucl.\ Instrum.\ Meth.\ A {\bf 344}, 319 (1994);  Nucl.\ Instrum.\ Meth.\ A {\bf 458}, 627 (2001).

\bibitem{bes3dect}
M. Ablikim {\it et al.} (BESIII Collaboration), Nucl.\ Instrum.\ Meth.\ A {\bf 614}, 345 (2010).

%\bibitem{Berger2010}
%N. Berger {\it et al.}, Chin. Phys. {\bf C34}, 1779 (2010).

\bibitem{Agostinelli:2002hh}
 S.~Agostinelli {\it et al.}  (GEANT4 Collaboration),
 %``GEANT4: A Simulation toolkit,''
 Nucl.\ Instrum.\ Meth.\ A {\bf 506}, 250 (2003).

\bibitem{kkmc2000}
 S.~Jadach, B.~F.~L.~Ward and Z.~Was,
  %``The Precision Monte Carlo event generator K K for two fermion final states in e+ e- collisions,''
 Comput.\ Phys.\ Commun.\  {\bf 130}, 260 (2000).

\bibitem{kkmc2001}
S.~Jadach, B.~F.~L.~Ward and Z.~Was,
  %``Coherent exclusive exponentiation for precision Monte Carlo calculations,''
Phys.\ Rev.\ D {\bf 63}, 113009 (2001).

%\cite{Ping:2008zz}
\bibitem{Ping2008}
R.~G.~Ping {\it et al.},
Chin. Phys. C {\bf 32}, 599 (2008).

\bibitem{pdg2012}
J.~Beringer {\it et al.}  (Particle Data Group),
%``Review of Particle Physics (RPP),''
Phys.\ Rev.\ D {\bf 86}, 010001 (2012).

\bibitem{Lundcharm}
%J.~C.~Chen {\it et al.},
J.~C.~Chen, G.~S.~Huang, X.~R.~Qi, D.~H.~Zhang and Y.~S.~Zhu,
Phys.\ Rev.\ D {\bf 62}, 034003 (2000).

\bibitem{GPUPWA}
%\bibitem{Berger:2010zza}
N.~Berger, B.~J.~Liu and J.~K.~Wang,
 %``Partial wave analysis using graphics processing units,''
 J.\ Phys.\ Conf.\ Ser.\  {\bf 219}, 042031 (2010).

\bibitem{Zou:2002ar}
B.~S.~Zou and D.~V.~Bugg,
%``Covariant tensor formalism for partial wave analyses of psi decay to mesons,''
Eur.\ Phys.\ J.\ A {\bf 16}, 537 (2003).

\bibitem{tensorphases}
J. G. Korner {\it et al.},
Phys. Lett. B {\bf 120}, 444 (1983).

\bibitem{Dymov:1998zu}
%S.~N.~Dymov, V.~S.~Kurbatov, I.~N.~Silin and S.~V.~Yaschenko,
S.~N.~Dymov {\it et al.},
  %``Constrained minimization in C++ environment,''
Nucl.\ Instrum.\ Meth.\ A {\bf 440}, 431 (2000).

\bibitem{22states}
We tested the following mesons listed in PDG 2012: $f_{2}(1270)$, $f_{0}(1370)$, $f_{2}(1430)$, $f_{0}(1500)$, $f_{2}^{'}(1525)$, $f_{2}(1565)$, $f_{2}(1640)$, $f_{0}(1710)$, $f_{2}(1810)$, $f_{2}(1910)$, $f_{2}(1950)$, $f_{2}(2010)$, $f_{0}(2020)$, $f_{4}(2050)$, $f_{0}(2100)$,  $f_{2}(2150)$, $f_{0}(2200)$, $f_{J}(2220)$, $f_{2}(2300)$, $f_{4}(2300)$, $f_{0}(2330)$, $f_{2}(2340)$.

\bibitem{bes2gammakk}
J.~Z.~Bai {\it et al.}  (BES Collaboration),
%``Partial wave analyses of J / psi ---> gamma K+ K- and gamma K0(S) K0(S),''
Phys.\ Rev.\ D {\bf 68}, 052003 (2003).

\bibitem{bes2gammapipi}
M.~Ablikim {\it et al.} (BES Collaboration),
%``Partial wave analyses of J/psi ---> gamma pi+ pi- and gamma pi0 pi0,''
Phys.\ Lett.\ B {\bf 642}, 441 (2006).

\bibitem{bes1gamma4pi}
J.~Z.~ Bai {\it et al.} (BES Collaboration),
%``Partial wave analysis of J / psi to gamma (pi+ pi- pi+ pi-),''
Phys.\ Lett.\ B {\bf 472}, 207 (2000).

\bibitem{Anisovich:1999fd}
A.~V.~Anisovich {\it et al.},
%``Study of the process anti-p p --> eta eta pi0 from 1350-MeV/c to 1940-MeV/c,''
Phys.\ Lett.\ B {\bf 449}, 145 (1999).

\bibitem{Anisovich:2000ut}
A.~V.~Anisovich {\it et al.},
%``I = 0 C = +1 mesons from 1920 to 2410 MeV,''
Phys.\ Lett.\ B {\bf 491}, 47 (2000).

\bibitem{Anisovich:2000ae}
A.~V.~Anisovich {\it et al.},
%``Partial wave analysis of anti-p p --> pi- pi+, pi0 pi0, eta eta and eta eta',''
Nucl.\ Phys.\ A {\bf 662}, 319 (2000).

\bibitem{bes2phipipi}
M.~Ablikim {\it et al.} (BES Collaboration),
Phys.\ Lett.\ B {\bf 607}, 243 (2005).

\bibitem{Baltrusaitis:1985pu}
R.~M.~Baltrusaitis {\it et al.}  (MARKIII Collaboration),
%``Observation of a Narrow K anti-K State in J/psi Radiative Decays,''
Phys.\ Rev.\ Lett.\  {\bf 56}, 107 (1986).

\bibitem{Bai:1998tx}
J.~Z.~Bai {\it et al.}  (BES Collaboration),
%``Experimental study of J / psi radiative decay to pi0 pi0,''
Phys.\ Rev.\ Lett.\  {\bf 81}, 1179 (1998).

\bibitem{depbw}
J.~H.~Kuhn, A.~Santamaria,
Z. Phys. C {\bf 48}, 445 (1990).

\bibitem{Ablikim:2011kv}
M.~Ablikim {\it et al.}  (BESIII Collaboration),
 %``Study of $\chi_{cJ}$ radiative decays into a vector meson,''
Phys.\ Rev.\ D {\bf 83}, 112005 (2011).

\bibitem{Ablikim:2010zn}
M.~Ablikim {\it et al.} (BESIII Collaboration),
Phys.\ Rev.\ D {\bf 81}, 052005 (2010).

\bibitem{Gui:2012gx}
L.~C.~Gui {\it et al.},
%Y.~Chen, G.~Li, C.~Liu, Y.~B.~Liu, J.~P.~Ma, Y.~B.~Yang and J.~B.~Zhang,
%``Scalar glueball in radiative $J/\psi$ decay on lattice,''
arXiv:1206.0125.

\end{thebibliography}
\end{document}